\documentclass{IEEEtran}
\usepackage{cite}
\usepackage{amsmath,amssymb,amsfonts}
\usepackage{algorithmic}
\usepackage{graphicx}
\usepackage{makecell}
\usepackage{caption}
\usepackage{subcaption}
\usepackage{dblfloatfix} 
\usepackage{multirow}
\usepackage{svg}
\usepackage{lipsum}
\usepackage{siunitx}
\usepackage{multirow}
\usepackage{cuted}
\usepackage{tabularx} 
\usepackage{booktabs}
\usepackage{multirow} 
\usepackage{svg}

\usepackage{amsmath}

\makeatletter

\newcommand{\Rmnum}[1]{\expandafter\@slowromancap\romannumeral #1@}
\makeatother

\def\BibTeX{{\rm B\kern-.05em{\sc i\kern-.025em b}\kern-.08em
    T\kern-.1667em\lower.7ex\hbox{E}\kern-.125emX}}
\begin{document}
\title{High-Precision Hybrid FA-PSO Based Inversion of Building Material Parameters for Fundamental Wireless Performance Evaluation}
\author{Zhuowei Li, \IEEEmembership{Member, IEEE}, Yalei Zhu, Hanqing Zhang, Sui Li, Meng Chen, Tong Zhang,  \\Zi-Yang Wu,  \IEEEmembership{Member, IEEE},  Dan Yang, Songjiang Yang, \IEEEmembership{Member, IEEE}, Jiliang Zhang, \IEEEmembership{Senior Member, IEEE}
\thanks{
Corresponding Author: Jiliang Zhang (zhangjiliang1@mail.neu.edu.cn)

Z. Li, Y. Zhu, H. Zhang, Z.-Y. Wu, D. Yang, and  J Zhang are with the College of Information Science and Engineering, Northeastern University, China, 110819.

S. Li is with the College of Jangho Architecture, Northeastern University, China, 110169.

M. Chen and T. Zhang are with the School of Resources and Civil Engineering, Northeastern University, China, 110819.

S. Yang is with the Purple Mountain Laboratories, China, 211111.

Z. Li and Y. Zhu contributed equally to this paper.

{This work was supported by the National Key R\&D Program of China under Grant 2025YFE0122200, the National Natural Science Foundation of China (NSFC) under Grant 62573096 and 62401644, the Postdoctoral Fellowship Program and China Postdoctoral Science Foundation under Grant Number BX20250343, the LiaoNing Revitalization Talents Program under Grant XLYC2403116, the Opening Fund of Liaoning Key Laboratory of Urban and Architectural Digital Technology UADT2024A05, and the Basic Scientific Research Fund of The State Key Laboratory of Synthetical Automation for Process Industries.} }
}

\maketitle

\begin{abstract}

 In this paper, we propose an inversion method based on the firefly particle swarm optimization (FA-PSO) algorithm to estimate the permittivity, conductivity, and thickness of building materials using the free-space method. To improve convergence efficiency and robustness, an adaptive firefly algorithm (FA) is employed to systematically optimize the hyperparameters of the particle swarm optimization (PSO). By optimizing the parameters of the Gaussian distribution used for population initialization, the accuracy of parameter estimation is gradually improved. Furthermore, we derive the Cramér–Rao lower bound (CRLB) for the permittivity, conductivity, and thickness under a complex Gaussian noise model, which serves as a theoretical benchmark for evaluating the estimation accuracy of the FA-PSO algorithm. Numerical results indicate that for relatively thin materials, the estimation accuracy of the proposed method approaches this theoretical lower bound, confirming the effectiveness of the inversion framework. This study accurately extracts the electromagnetic properties of building materials, providing strong support for evaluating their wireless performance.

%The electromagnetic (EM) wave propagation indoors would be affected by the EM and physical properties of the building material, e.g., its relative permittivity and thickness. Accurately extracting electromagnetic parameters is crucial for evaluating the wireless performance of building materials. In this paper, we build a measurement platform and obtain the S21 parameters of building materials in an indoor environment. This study proposes an inversion method based on a firefly-assisted particle swarm optimization (FA-PSO) algorithm. To improve convergence efficiency and robustness, the hyperparameters of particle swarm optimization (PSO) are systematically optimized using an adaptive firefly algorithm (FA). Additionally, the accuracy of parameter extraction is progressively enhanced by optimizing the parameters of the Gaussian distribution used for population initialization. We further derive the Cramér–Rao lower bound (CRLB) for the estimated parameters as a theoretical benchmark for estimator optimality. Numerical results show that for relatively thin materials, the proposed method achieves estimation accuracy close to this theoretical bound, confirming the effectiveness of the inversion framework. This study accurately extracts the electromagnetic properties of building materials, providing effective support for evaluating their wireless performance.
\end{abstract}


\section{Introduction}
\label{sec:introduction}
\IEEEPARstart{A}{s} wireless communication technology continues to evolve from 5G to 6G, indoor wireless coverage quality is important [1]. Buildings are the primary physical environment for wireless signal propagation. Building structural layout and characteristics of the materials used significantly affect the propagation path, penetration loss, and multipath effects of electromagnetic waves, severely limiting indoor wireless signal quality [2], [3]. In particular, the use of inappropriate building materials can seriously impede signal penetration, leading to a decline in indoor communication quality and even the formation of signal dead zones.

In search of building materials with wireless friendly properties, Zhang \textit{et al}. proposed the spatially averaged capacity, which was specifically designed to quantify the impact of building materials on wireless penetration performance and depends primarily on their permittivity, conductivity, and thickness. It lays the theoretical foundation for evaluating the wireless performance of buildings [4], [5]. However, in order to accurately calculate this core indicator, the critical technical challenge is how to extract the complex electromagnetic characteristics of building materials with high precision, especially their permittivity, conductivity, and thickness.

Currently, the main techniques for measuring the complex permittivity of materials included the resonant cavity methods [6]-[8], the waveguide penetration/reflection methods [9], [10], and the coaxial probe methods [11]. The cavity resonant method stands out as a highly precise and extensively researched technique. Nevertheless, its reliance on meticulous specimen preparation for accuracy and its operation within a narrow frequency range are notable limitations [12]. In waveguide measurements, the sample must fully fill the cross-section of the waveguide, as any gaps introduce significant errors [13], [14]. Additionally, when the sample has a finite thickness, the phase of the measured scattering parameters becomes ambiguous, leading to multiple possible solutions for the dielectric constant. In contrast, the free-space method is a non-contact measurement technique that requires no complex sample preparation or strict dimensional constraints. At the same time, it could perform sweep measurements over a very wide frequency range [15]-[18]. Given these advantages, the free-space method was the ideal choice for measuring the complex permittivity of solid samples like building materials.

In homogeneous dielectric materials, the reflection coefficient can approach zero at certain frequencies [19], a phenomenon typically arising from destructive interference between the incident and reflected waves. Although such zero-reflection behaviour has been observed in previous measurements, it has rarely been discussed in detail, and the extremely weak signal at the zero-reflection point makes it difficult to further retrieve the electromagnetic parameters of the material. Therefore, accurately extracting the intrinsic electromagnetic parameters of materials from measured received power was a highly nonlinear and ill-posed inverse problem, which constituted a critical bottleneck for the practical application of the free-space method. Although intelligent algorithms such as particle swarm optimization (PSO) partially reduced the multi-value solution issue caused by phase ambiguity in traditional methods like Nicolson-Ross-Weir, their performance relied on parameter settings and they were prone to falling into local optima when dealing with high-dimensional, nonlinear inverse problems, compromising the global optimality and accuracy of the solution [20], [21]. Moreover, existing inversion algorithms primarily focused on extracting the complex permittivity, while seldom treating material thickness as a key parameter to be simultaneously inverted [22], [23], which limited their applicability in scenarios that require concurrent characterization of both electromagnetic and geometric properties. Furthermore, most current studies emphasized performance improvement through comparative evaluations of different algorithms, but lacked in depth exploration of the theoretical performance limits of the inversion methods themselves [24], [25], leaving their applicability and reliability insufficiently addressed.

In this paper, we proposes an electromagnetic parameter inversion method for building materials based on a firefly algorithm and particle swarm optimization (FA‑PSO) algorithm. The penetration coefficient and reflection coefficient were experimentally measured using a vector network analyzer (VNA) in the frequency range of 20–35 GHz. The complex permittivity was extracted using the FA-PSO algorithm and compared with the traditional PSO algorithm to verify the accuracy of the estimation inversion. The main contributions of this paper are summarized as follows:
\begin{enumerate}

\item A parameter inversion framework is proposed to extract the permittivity, conductivity, and thickness of building materials by the realistic measurements. The framework consists of $S_{21}$ measurement, penetration coefficient calculation from the measured $S_{21}$, data preprocessing, and parameter inversion. This framework establishes a complete workflow for inverting a material’s electromagnetic parameters from $S_{21}$.

%Specifically, penetration coefficients are derived from measured $S_{21}$ and preprocessed to mitigate indoor noise and interference prior to inversion of permittivity, conductivity, and thickness.

\item A FA-PSO algorithm is proposed to extract a material’s electromagnetic and physical properties, including permittivity, conductivity, and thickness. To improve convergence efficiency and robustness, this study systematically optimizes the PSO hyper parameters using an adaptive firefly algorithm (FA). By optimizing the parameters of the Gaussian distribution used for initializing the population in FA-PSO, the accuracy of parameter extraction by this algorithm is progressively improved.

\item The reflection coefficient for the homogeneous medium closed to zero within certain frequency bands was observed in measurements because of destructive interference arising from the phase difference between the reflections at the front and rear surfaces. Since the high-precision global optimization capability of the proposed algorithm, the accurate electromagnetic parameters of the material can be extracted from very low measured reflection coefficients to infer the material class.

\item The Cramér–Rao lower bound (CRLB) for the parameters, including the electromagnetic parameters and thickness, is derived to vaildate whether the estimator of the proposed algorithm achieves optimal performance. The results indicate that, for relatively thin materials, the proposed method can achieve an estimation accuracy close to this theoretical lower bound.

%\item The results indicate that, for relatively thin materials, the proposed method can achieve an estimation accuracy close to this theoretical lower bound. Meanwhile, we reveal a statistically significant correlation between the inversion accuracy of the proposed algorithm and the material thickness, indicating a clear thickness dependent behavior.

\end{enumerate}

The remainder of this paper is organized as follows. Section \Rmnum{2} introduces the measuring instruments and operating methods. Section \Rmnum{3} explains the selection of optimization algorithms. In Section \Rmnum{4}, the measurement results are reported and discussed. Finally, Section \Rmnum{5} summarizes the conclusions.

\begin{figure}[!t]
    \begin{minipage}{1\linewidth}
        \centering
        % 第一行图片
        \begin{subfigure}{\textwidth}
            \centering
            \includegraphics[width=1\textwidth]{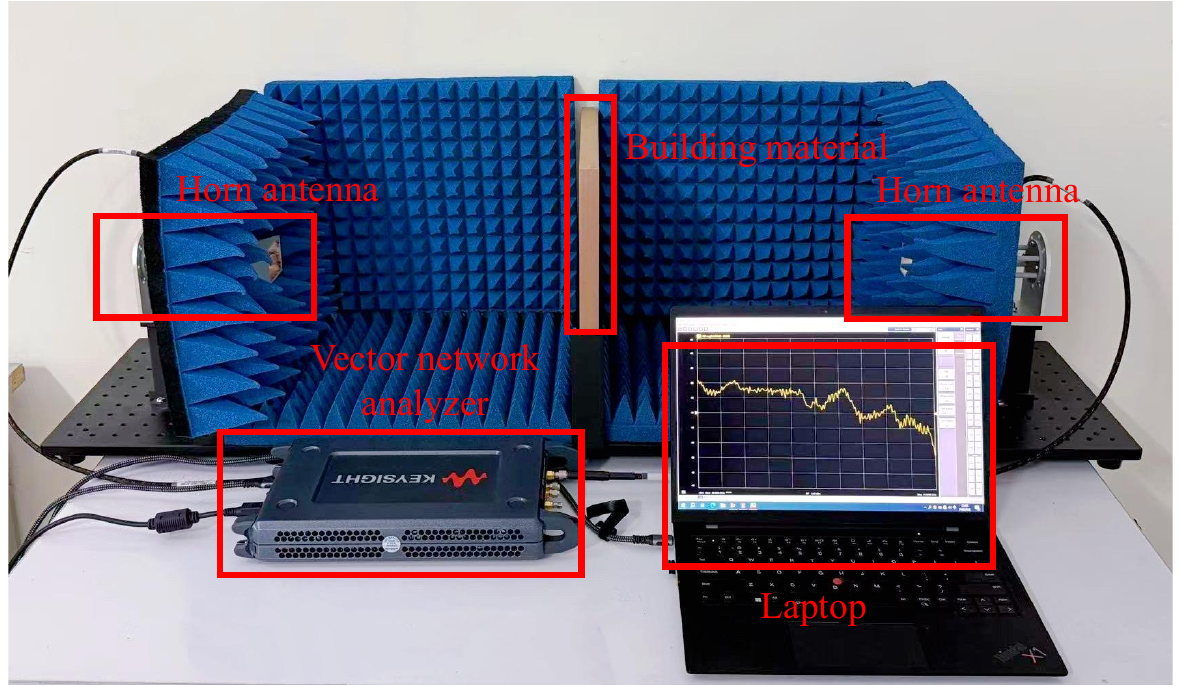}
            \caption{}
            %\label{fig:sub1}
        \end{subfigure}
        % 第二行图片
        \begin{subfigure}{\textwidth}
            \centering
            \includegraphics[width=1\textwidth]{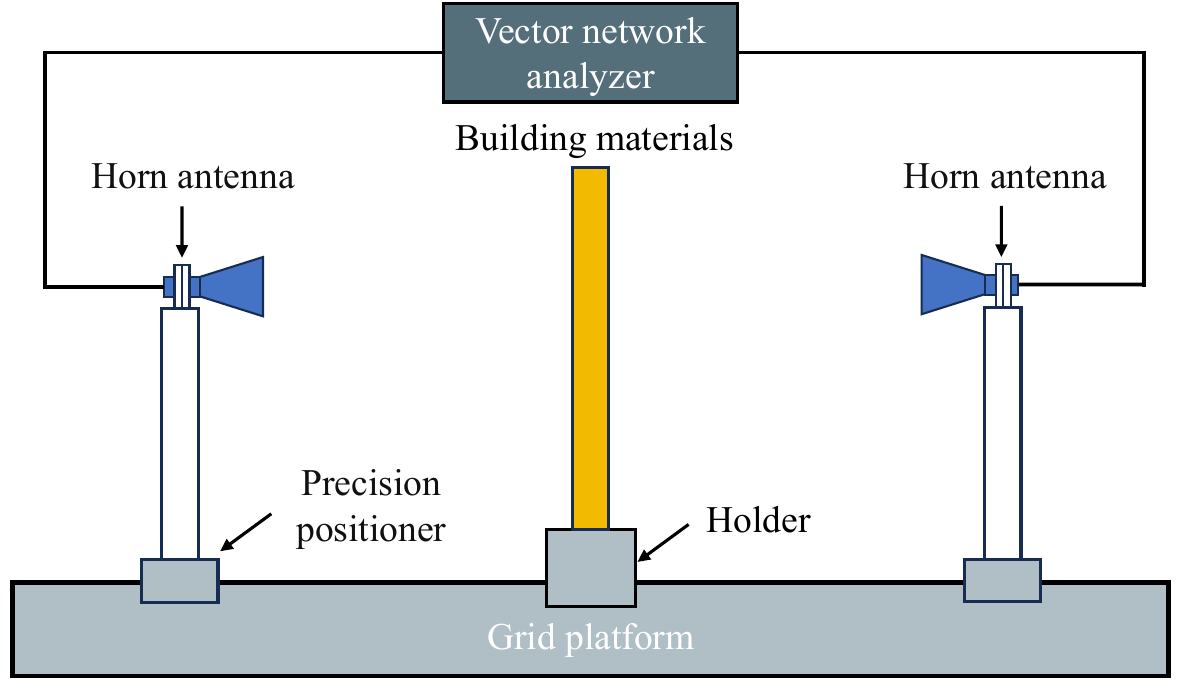}
            \caption{}
            %\label{fig:sub2}
        \end{subfigure}
    \end{minipage}
    \hfill
\caption{Measurement platform. (a) Realistic experimental setup. (b) Schematic diagram.}
\label{fig1}
\end{figure}

\begin{figure*}[!t]
    \centering % 使用 \centering 命令，比 \centerline 更好

    % --- 第一个子图 ---
    \begin{subfigure}[b]{0.7\columnwidth} % 设置第一个子图的宽度
        \centering
        \includegraphics[width=\linewidth]{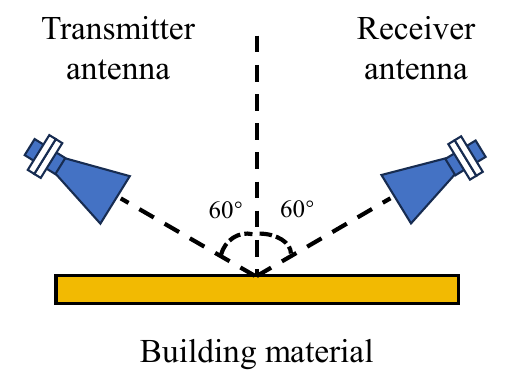} % <-- 替换成你的第一张图片文件名
        \caption{}
        \label{fig:6a} % 为第一个子图添加标签（可选）
    \end{subfigure}
    \quad % 这个命令会自动在两个子图之间添加空白，使它们两端对齐
    % --- 第二个子图 ---
    \begin{subfigure}[b]{0.7\columnwidth} % 设置第二个子图的宽度
        \centering
        \includegraphics[width=\linewidth]{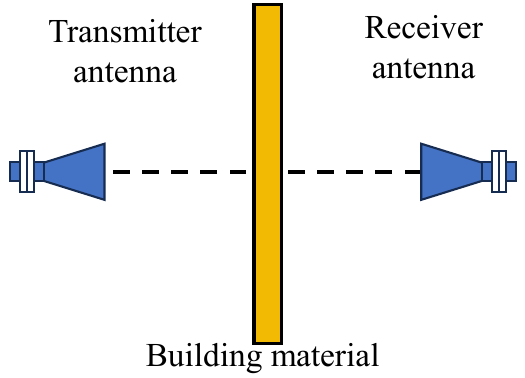} % <-- 替换成你的第二张图片文件名
        \caption{}
        \label{fig:6b} % 为第二个子图添加标签（可选）
    \end{subfigure}

    % --- 主标题 ---
    % (a) 和 (b) 是由 subfigure 自动生成的，主标题中不需要再写
    \caption{Schematic diagram of the measurement configurations. (a) Reflection. (b) Penetration. }
    \label{fig6} % 整个图片环境的标签
\end{figure*}

\section{Measurement Apparatus and Operation}

In this section, we measured the reflection and penetration coefficients of the material under test (MUT), including acrylic, rubber, and bakelite. The measurement was conducted in an indoor environment using a dual-antenna measurement system.  The experimental procedure is described in detail.

\subsection{Measurement Platform}

The measurement system is composed of a keysight P9377B VNA, two horn antennas, and a laptop. In this study, the VNA was set to scan 1093 sampling points in the 20-35~GHz frequency band, with a frequency resolution of 15 MHz. The VNA was calibrated by using the conventional two port short-open-load-through (SOLT) methodology, with coaxial cables and connectors used to disentangle the system's inherent frequency response from the measurements. The transmit power was set to 0 dBm and the intermediate frequency bandwidth was set to 1 kHz to balance the measurement time and the dynamic range of the system. To minimize systematic errors, the number of measurements was averaged over five repetitions per frequency point. The measurement configuration used in this study is shown in Fig. \ref{fig1}.

The transmitter (Tx) and receiver (Rx) antennas are a pair of horn antennas with a diameter of 50.5 mm × 50.5 mm and an operating frequency of 20 to 35 GHz. The antennas are mounted on a tripod at a height of 70 cm above the ground to meet far-field assumption, and the distance between both antennas and the MUT is set at 50 cm. The setup of reflection coefficient measurement is shown in Fig. 2 (a), in which the Rx antenna is placed at the specular reflected direction of Tx antenna referring to material slab surface. The penetration coefficients are measured based on the setup shown in Fig. 2 (b), in which the Tx and Rx antennas are placed on the opposite sides of the material slab. Microwave absorbing materials were employed around the measurement platform to mitigate multi-path component during the measurement.
%吸波材料is 布署ed around the measurement platform to eliminate multi-path component during the measurement.

\begin{figure}[!t] 
    \centering % 使用 \centering 命令，比 \centerline 更好

    % --- 第一个子图 --- 
    \begin{subfigure}[b]{0.29\columnwidth} 
        \centering
        \includegraphics[width=\linewidth]{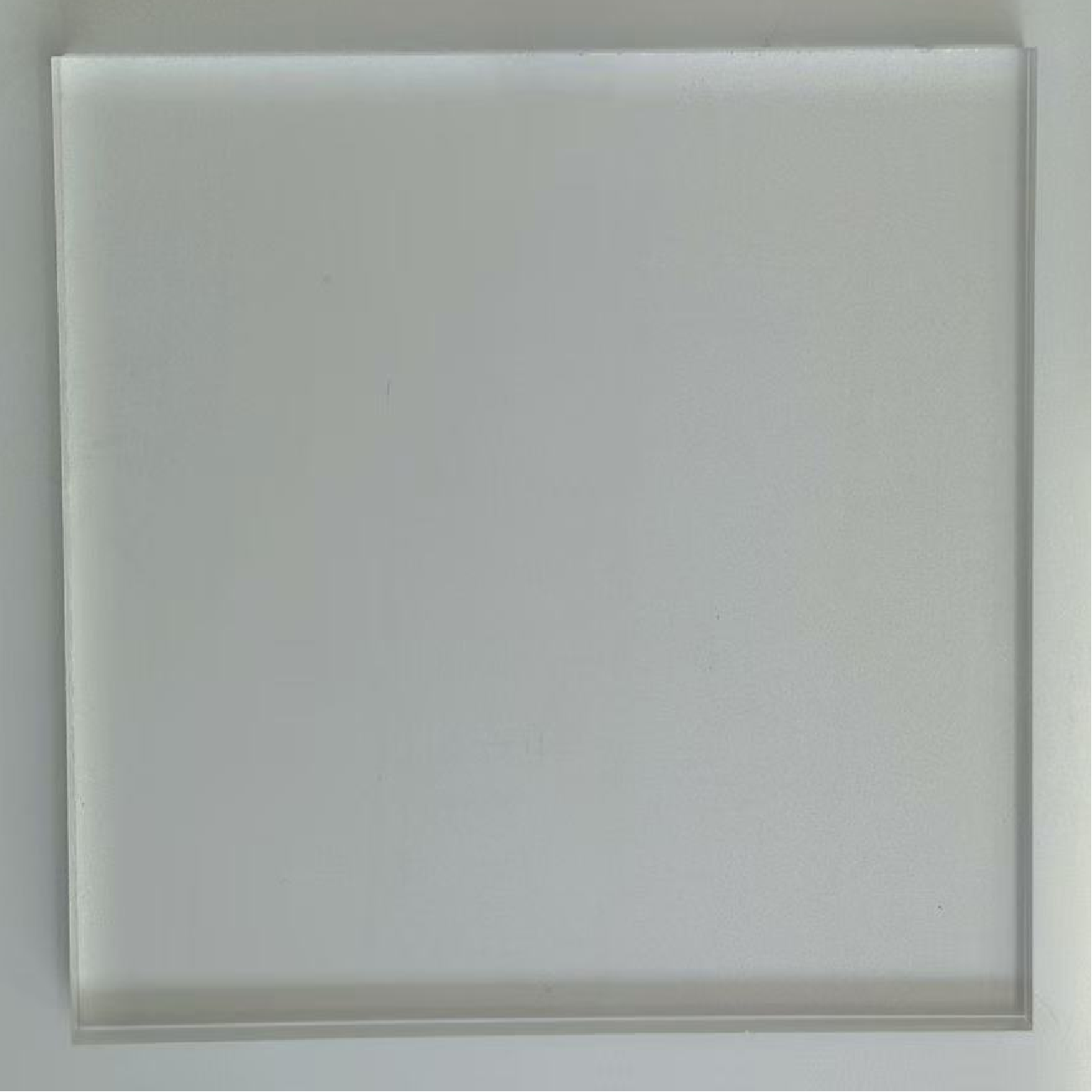}
        \caption{} 
        \label{fig:5a} 
    \end{subfigure} 
    \quad % 在图a和图b之间添加空白
    % --- 第二个子图 --- 
    \begin{subfigure}[b]{0.29\columnwidth}
        \centering
        \includegraphics[width=\linewidth]{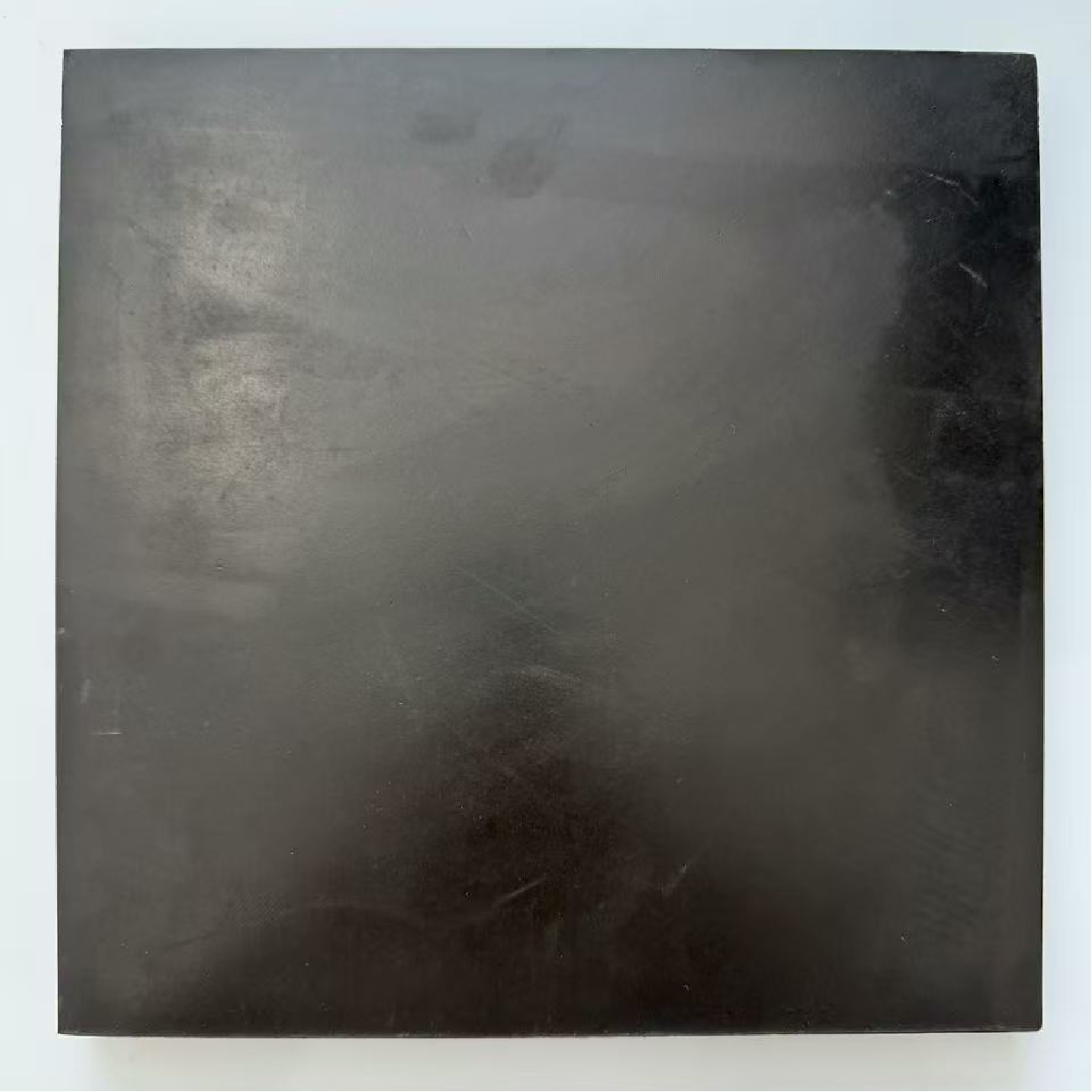}
        \caption{} 
        \label{fig:5b}
    \end{subfigure} 
    \quad % <-- 在图b和图c之间也加上这个！
    % --- 第三个子图 --- 
    \begin{subfigure}[b]{0.29\columnwidth}
        \centering
        \includegraphics[width=\linewidth]{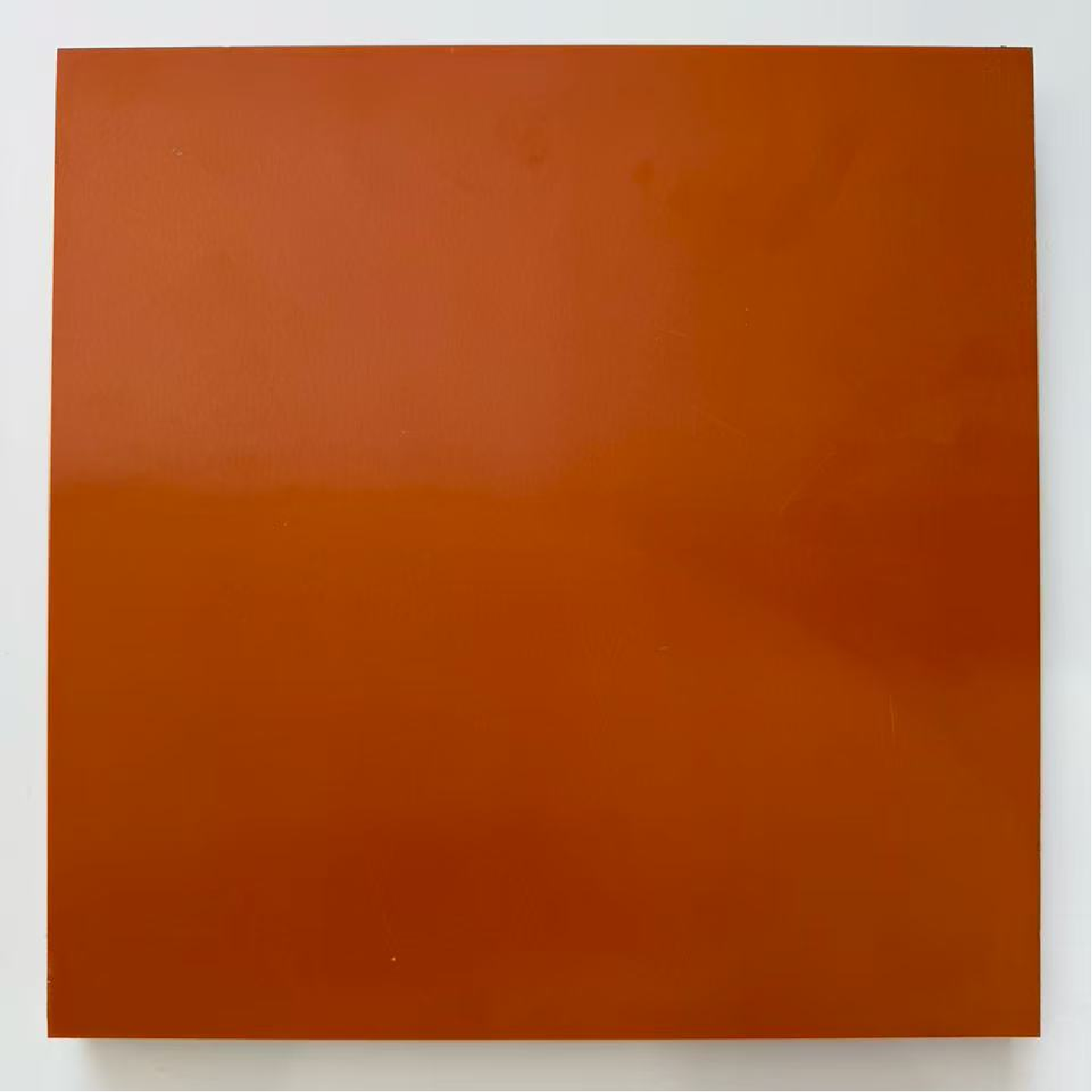}
        \caption{} 
        \label{fig:5c}
    \end{subfigure} 
    % <-- 这里末尾的 \hfill 已经不需要了，可以删除
    
    % --- 主标题 --- 
    \caption{Photographs of the three MUT samples. (a) Acrylic. (b) Rubber. (c) Bakelite. } 
    \label{fig5}
\end{figure}

\begin{figure}[!t] 
    \centering % 使用 \centering 命令，比 \centerline 更好

    % --- 第一个子图 --- 
    \begin{subfigure}[b]{0.29\columnwidth} 
        \centering
        \includegraphics[width=\linewidth]{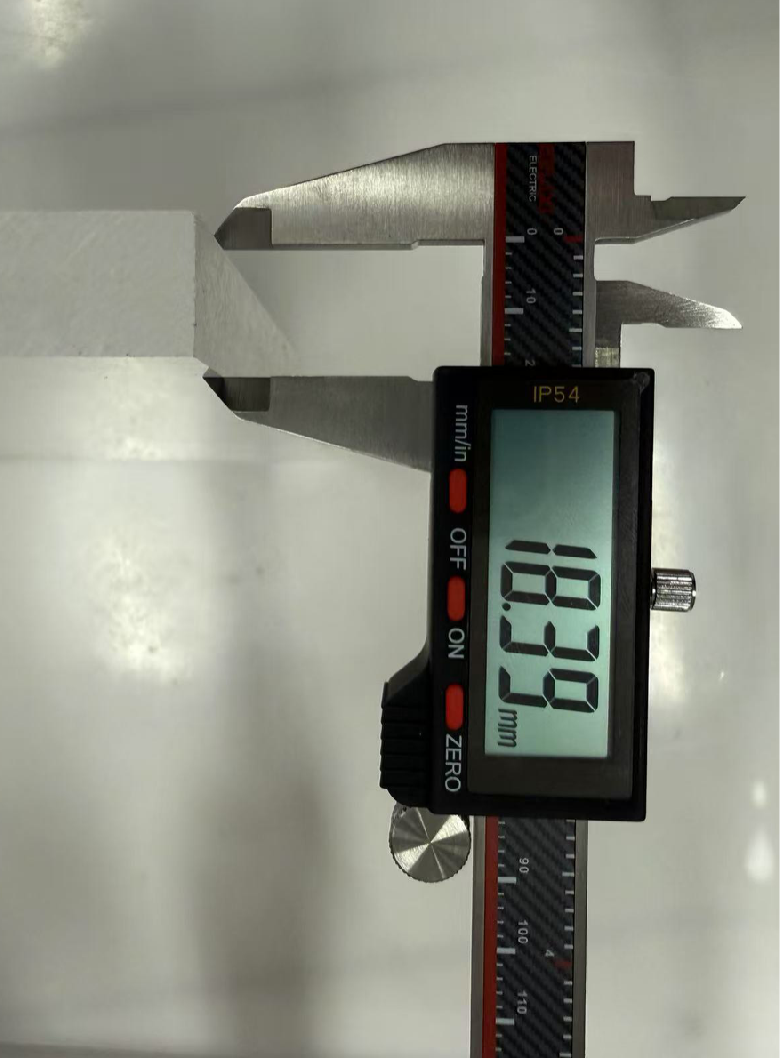}
        \caption{} 
        \label{fig:9a} 
    \end{subfigure} 
    \quad % 在图a和图b之间添加空白
    % --- 第二个子图 --- 
    \begin{subfigure}[b]{0.29\columnwidth}
        \centering
        \includegraphics[width=\linewidth]{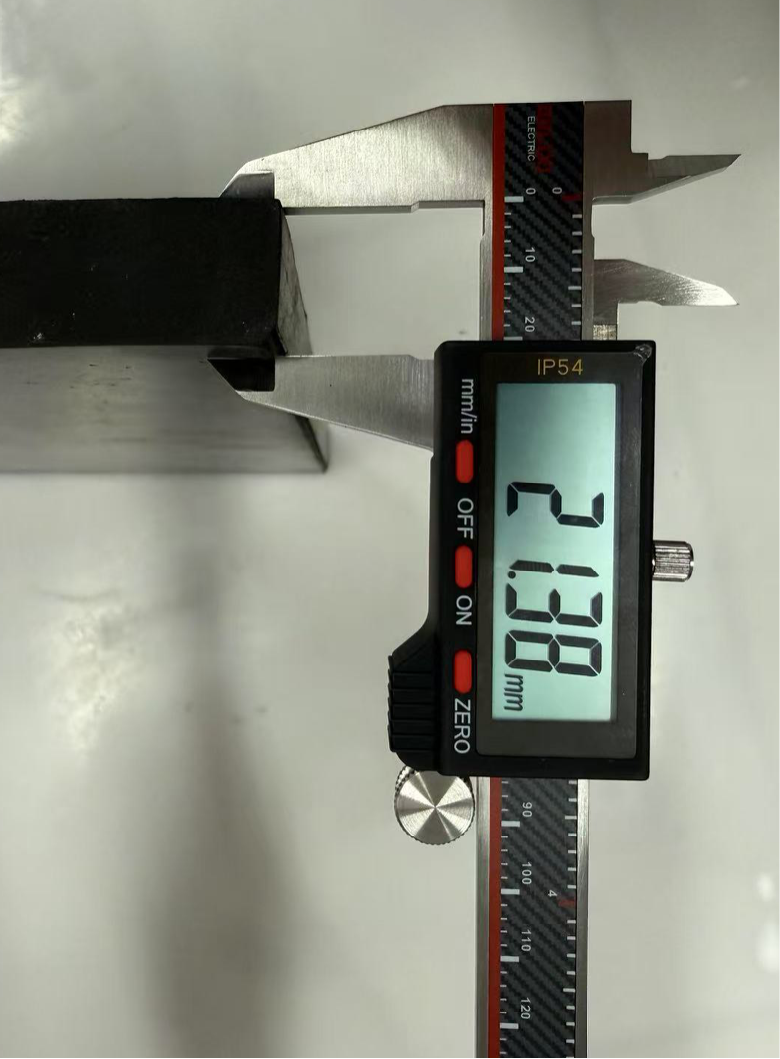}
        \caption{} 
        \label{fig:9b}
    \end{subfigure} 
    \quad % <-- 在图b和图c之间也加上这个！
    % --- 第三个子图 --- 
    \begin{subfigure}[b]{0.29\columnwidth}
        \centering
        \includegraphics[width=\linewidth]{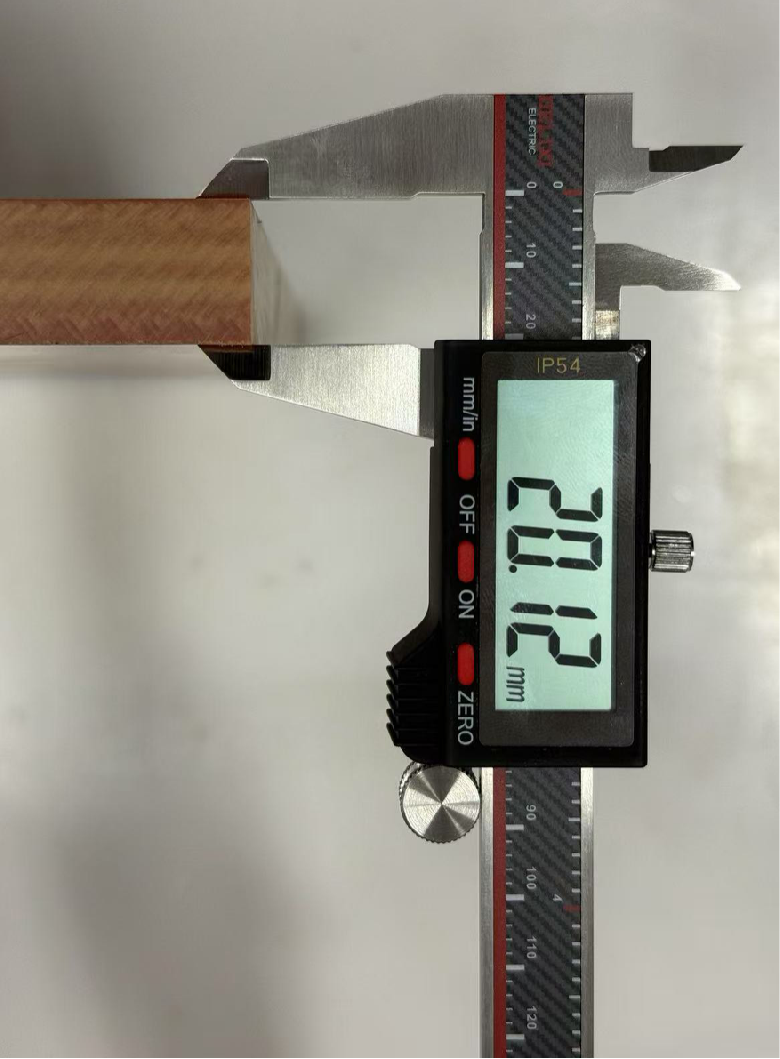}
        \caption{} 
        \label{fig:9c}
    \end{subfigure} 
    % <-- 这里末尾的 \hfill 已经不需要了，可以删除
    % --- 主标题 --- 
    \caption{Measured thickness values for the three MUT samples. (a) Acrylic. (b) Rubber. (c) Bakelite. } 
    \label{fig9}
\end{figure}

Extracting electromagnetic parameters for single-component materials is relatively simple. However, for multicomponent or inhomogeneous composites, parameter extraction remains a major challenge due to the complexity of their internal structure. Therefore, in order to verify the excellent performance of our algorithm, we selected three typical building materials commonly used in construction as research objects and measured the penetration coefficients of acrylic, rubber, and bakelite boards in the 20-35 GHz frequency band. MUT are shown in Fig. \ref{fig5} and Fig. \ref{fig9} shows the corresponding thickness measured simultaneously. These material samples are large enough to minimize diffraction from the edges. 

\subsection{Data Processing Procedures}

\begin{figure}[!t]
\centerline{\includegraphics[width=0.95\columnwidth]{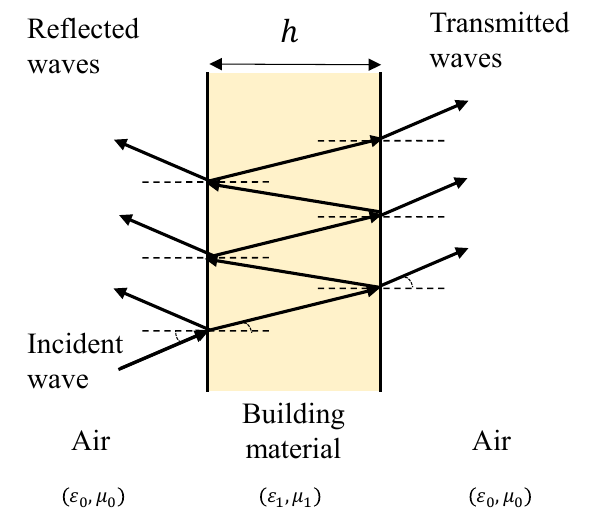}}
\caption{Penetration  and reflection of a plane wave for a single-layer slab.}
\label{fig2}
\end{figure}

To reduce indoor noise and multipath reflections and capture the true electromagnetic response of the material, we preprocessed the VNA data. First, we use the inverse Fourier transform to change the S parameter from the frequency domain to the time domain. Next, we used time-domain gating to keep the direct signal and remove system noises. Finally, we converted the gated signal back to the frequency domain for dielectric parameter inversion.

The method involves measuring the sample's penetration coefficient and calculating it from the theoretical model. The complex permittivity is then extracted by minimizing the root mean squared error (RMSE) between the measured and model-predicted vertically polarized penetration coefficients. RMSE is defined as

\begin{equation}
\text{RMSE} = \sqrt{ \frac{1}{\mathrm{N}} \sum_{f}^{f_\text{max}} \left| \mathit\Gamma^{\text{cal}}(f) - \mathit\Gamma^{\text{meas}}(f) \right|^2 },
\end{equation}
where $\mathit{\Gamma}^{\text{cal}}(f)$ and $\mathit{\Gamma}^{\text{meas}}(f)$ are the calculated and measured penetration coefficient, $\mathrm{N}$ is the number of frequency points, and the frequency f ranges from 20 to 35 GHz. The penetration coefficient is measured with an incident angle of 0°.

The penetration coefficient is obtained from the frequency response measured with the MUT, copper board as PEC, and air. To ensure that both measurements are conducted under identical conditions, the Tx and Rx antennas remain fixed, and only the MUT is replaced. The measured penetration and reflection coefficients are expressed as [26]
\begin{align}
T_{\mathrm{meas}}(f) &= \frac{S_{21}^{\mathrm{mat}}(f)}{S_{21}^{\mathrm{air}}(f)}, \\
\mathit\Gamma_{\mathrm{meas}}(f) &= \frac{S_{21}^{\mathrm{mat}}(f)}{S_{21}^{\mathrm{PEC}}(f)},
\end{align}
where ${S_{21}^{\mathrm{air}}(f)}$ is the frequency response of the Tx and Rx antennas in free space without the MUT, and ${S_{21}^{\mathrm{mat}}(f)}$ represents the frequency response of the material, while  ${S_{21}^{\mathrm{PEC}}(f)}$ is the frequency response measured with the copper board.

Fig. \ref{fig2} schematically illustrates the wave propagation process when a plane wave is incident on a single layer plate, including transmission within the medium and reflections generated at the interfaces. The penetration coefficient $\mathit{\Gamma}$ and reflection coefficient $T$ of a single flat material can be expressed as~[27]
\begin{align}
T = \frac{(1 - \mathit{\Gamma}'^2)e^{-j\beta h}}{1 - \mathit{\Gamma}'^2 e^{-2j\beta h}}, \\
\mathit{\Gamma} = \frac{\mathit{\Gamma}'(1 - e^{-2j\beta h})}{1 - \mathit{\Gamma}'^2 e^{-2j\beta h}},
\end{align}
where
\begin{equation}
\mathit{\Gamma}' = \frac{\cos\theta_i - \sqrt{\varepsilon_r - \sin^2\theta_i}}{\cos\theta_i + \sqrt{\varepsilon_r - \sin^2\theta_i}},
\end{equation}
\begin{equation}
\beta = \frac{2\pi}{\lambda} \sqrt{\varepsilon_r - \sin^2 \theta_i},
\end{equation}
\begin{equation}
\varepsilon_r = \varepsilon' - j\varepsilon'' = \varepsilon' - j\frac{\sigma}{\omega\varepsilon_0},
\end{equation}
\begin{equation}
\sigma = cf^{d},
\end{equation}
$\theta_i$ is at the angle between the ray and the normal, $h$ is the thickness of the material, $\beta$ is the propagation constant, $\lambda$ is the free-space wavelength, $\varepsilon_r$ is the complex permittivity, $\varepsilon'$ is the permittivity, $\sigma$ is the conductivity, $\omega$ is the angular frequency, $\varepsilon_0$ is the vacuum dielectric constant, $c$ and $d$ are determined as functions of $\sigma$, $\mathit{\Gamma}'$ is Fresnl reflectance coefficient of pure~polarization.

\section{FA-PSO Algorithm}
To slove the above non-linear optimization problem, this section presents a hybrid optimization algorithm. Standard PSO depends strongly on its parameter settings. It can also converge too early and get stuck in a local optimum for complex problems. Therefore, we propose a FA-PSO algorithm. In this method, the FA adaptively tunes the key PSO parameters. The core principle of this method is to take advantage of the global search capability of the FA to dynamically adjust the key control parameters of the PSO, namely the inertia weight $w$, the cognitive learning factor $c_1$ and social learning factor $c_2$. This approach overcomes the deficiencies of the conventional PSO algorithm, such as its susceptibility to local optima and slow convergence, which stem from its dependence on fixed values of $w$, $c_1$, and $c_2$.

\begin{figure}[!t]
\centerline{\includegraphics[width=1\columnwidth]{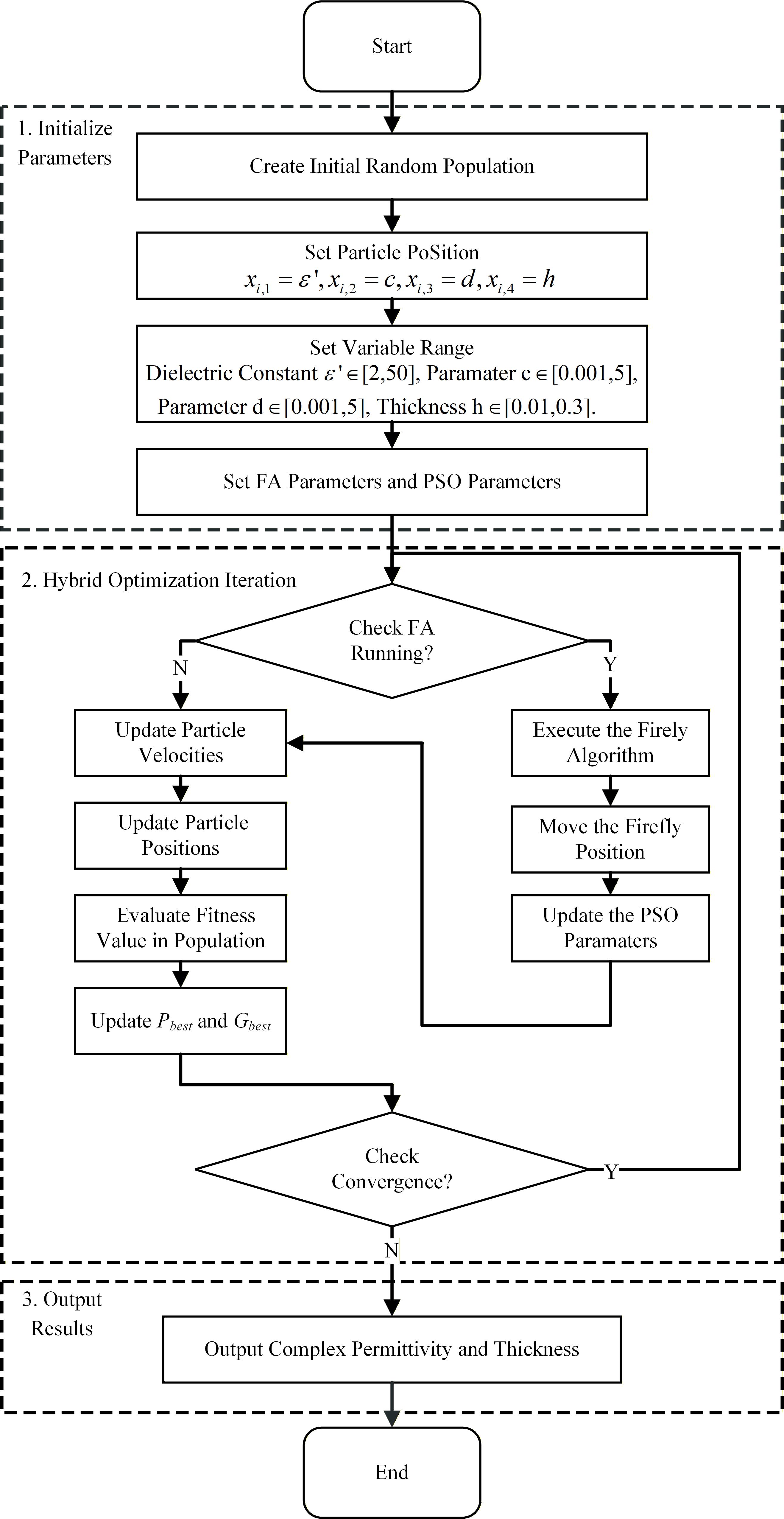}}
\caption{Overall procedure of complex permittivity extraction based on FA-PSO algorithm.}
\label{Fig6}
\end{figure}

\begin{figure}[!t]
\centerline{\includegraphics[width=1\columnwidth]{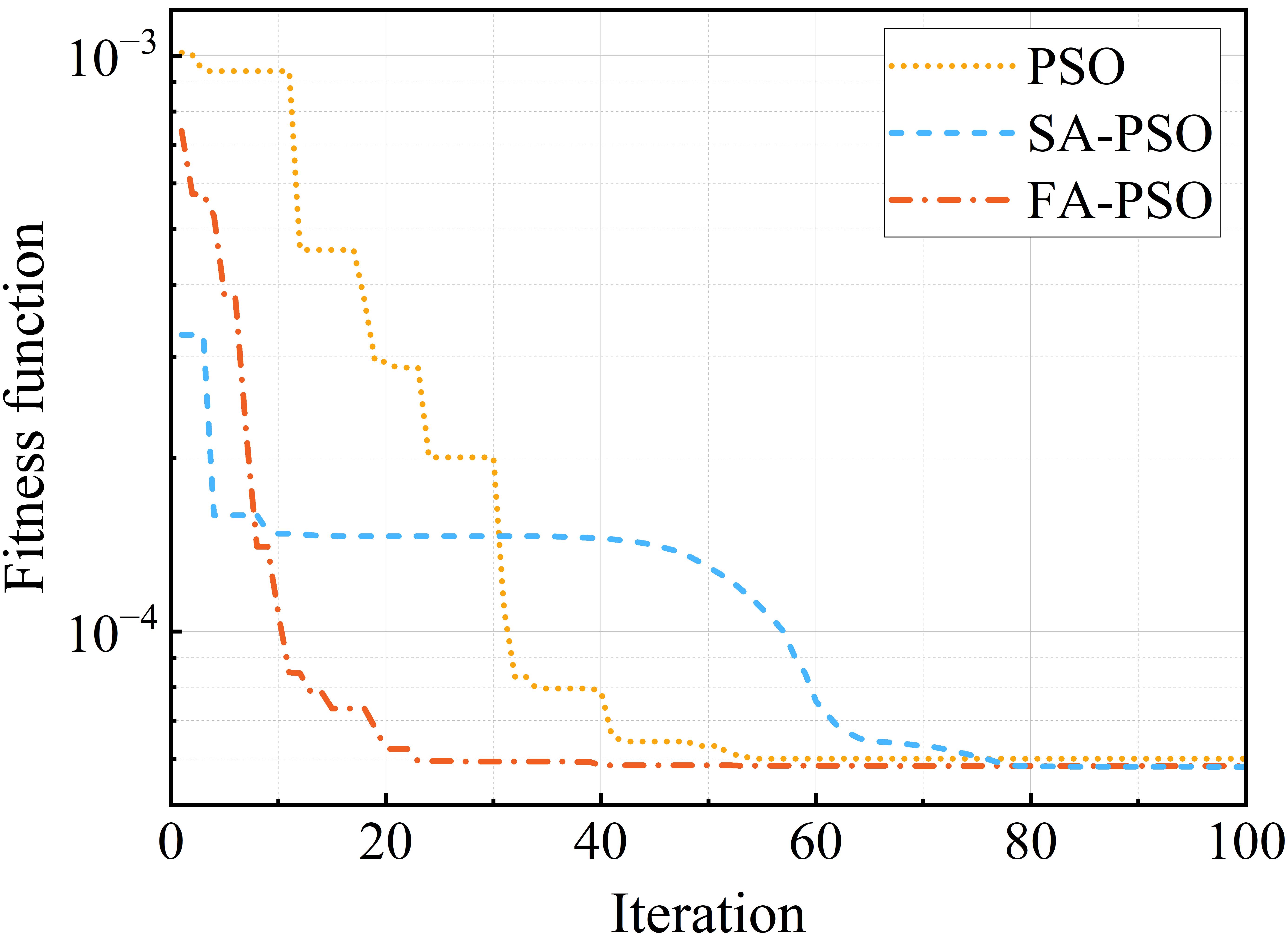}}
\caption{Iteration process of PET based on FA-PSO algorithm.}
\label{Fig7}
\end{figure}

\subsection{Particle Swarm Optimization}
PSO is a swarm intelligence optimization algorithm inspired by the foraging behavior of birds [28]. The algorithm initializes a population of randomly distributed particles, each of which represents a potential solution to the optimization problem. In each iteration, particles update their velocities and positions based on their individual best position $P_{best}$ and the global best position $G_{best}$ found by the entire population. The update equations are express as [29]
\begin{align}
\nonumber
v_{k,l}(t+1) &= \omega v_{k,l}(t) + c_1  r_1 (p_{id} -  x_{k,l}(t)) \\
             &\quad + c_2  r_2  (p_{gd} - x_{k,l}(t)),
\end{align}
\begin{align}
x_{k,l}(t+1) &= x_{k,l}(t) + v_{k,l}(t),
\end{align}
where $k$ is the particle index and $l$ is the dimension index, $r_1$ and $r_2$ are random numbers uniformly distributed between 0 and 1, $c_1$ and $c_2$ represent the cognitive and social learning coefficients, which are initialized to 1.5 and 2.0, respectively, in this algorithm. The inertia weight $\omega$, is set to 0.5.

\begin{figure*}[t]

    \vspace{2pt}
    % --- 使用 equation + split 环境来实现单编号、多行对齐 ---
    \begin{equation}  % <-- 从 align 改为 equation
    \tag{18}
    \begin{split}     % <-- 在内部嵌套一个 split 环境
    \frac{\partial^2(\ln \ell)}{\partial\phi_i \partial\phi_j} = \frac{1}{\kappa^2} \sum \operatorname{Re}\left[ \frac{\partial^2 T(\phi; n)^*}{\partial\phi_i \partial\phi_j} (y[n] - T(\phi; n)) - \left(\frac{\partial T(\phi; n)}{\partial\phi_i}\right)^* \left(\frac{\partial T(\phi; n)}{\partial\phi_j}\right) \right].
    \end{split}       % <-- 结束 split
    \end{equation}    % <-- 结束 equation
    \hrulefill
\end{figure*}

The search space of this algorithm is defined by four parameters including the permittivity, the conductivity parameters $c$ and $d$, and the material thickness. To further enhance both the convergence efficiency and solution quality, a hierarchical hybrid strategy is adopted during the initialization phase. In this approach, the initial particle population is divided into three subsets, each serving a distinct purpose:
\begin{enumerate}

\item Random generation within the solution space to ensure population diversity.
\item Perturbation based on empirical values derived from prior knowledge to accelerate early convergence. 
\item Distribution around the center of the solution space to explore the central region. 
\end{enumerate}

\subsection{FA-PSO Algorithm}

The FA is an intelligent heuristic algorithm, which draws its inspiration from the flashing and mutual attraction behaviors of fireflies. The core principle of FA is that less bright fireflies are attracted to and move towards brighter ones.
The attractiveness $\Lambda$ of firefly $i$ toward firefly $j$ can be expressed as
\begin{equation}
\tag{12}
\Lambda(r) = \Lambda_0 \exp(-\gamma r^2),
\end{equation}
where $\Lambda{_0}$ is the maximum attractiveness when $r$ = 0, $\gamma$ is the light absorption coefficient, and $r$ represents the Euclidean distance between the two firefiles.

When firefly $o$ is attracted to a brighter firefly $p$, its position is expressed as [30]
\begin{equation}
\quad x_o^{t+1} = x_o^t + \Lambda(r)(x_p^t - x_o^t) + \alpha(\zeta - 0.5),
\end{equation}
where $\alpha$ is the randomization parameter, $\zeta$ is a uniformly distributed random number between 0 and 1. The second term governs the attractive movement, and the third term is a random perturbation that facilitates exploration.

The outer loop passes its best firefly to the inner-level PSO. This firefly holds the optimal parameters $[w, c_1, c_2]$. This adaptive strategy also controls computational cost. The overall procedures of estimation are shown in Fig. \ref{Fig6}. The values of the (1) during the iterative optimization of the poly ethylene terephthalate (PET) are shown in Fig. \ref{Fig7}, indicating that the FA-PSO algorithm achieves better convergence performance over the iteration process.

This paper proposes a Gaussian distribution-based initialization strategy to replace the uniform distribution. The Gaussian distribution incorporates domain-specific prior information into the optimization process: its mean $\mu$ may be set as the optimal prior estimate for parameters, while the standard deviation $\tau$ reflects the confidence level of this prior. A low $\tau$ concentrates the search when priors are reliable, whereas a high $\tau$ ensures diversity and exploration. Adapting flexibly $\mu$ and $\tau$, the Gaussian distribution integrates domain knowledge into the optimization process, facilitating a transition from random guesswork to strategic planning. 

\subsection{Signal Models and the CRLB Derivation}

\begin{figure*}[b!]
 \hrulefill 
    \vspace{2pt}
    % --- 使用 equation + split 环境来实现单编号、多行对齐 ---
    \begin{equation}  
     \tag{26}
    \begin{split}     % <-- 在内部嵌套一个 split 环境
     \frac{\partial^2 \ell(\boldsymbol{\theta})}{\partial \theta_i^2} \approx \frac{-\ell(\boldsymbol{\theta} + 2h_i \mathbf{e}_i) + 16\ell(\boldsymbol{\theta} + h_i \mathbf{e}_i) - 30\ell(\boldsymbol{\theta}) + 16\ell(\boldsymbol{\theta} - h_i \mathbf{e}_i) - \ell(\boldsymbol{\theta} - 2h_i \mathbf{e}_i)}{12h_i^2}.
    \end{split}       % <-- 结束 split
    \end{equation}    % <-- 结束 equation
\end{figure*}

The CRLB for the parameters, including the electromagnetic parameters and thickness, is derived to assess whether the estimator of the proposed algorithm achieves optimal performance. In the experimental environment of this study, absorptive materials were employed to suppress primary multipath interference. Consequently, we formulate the following standard statistical hypothesis for the noise process $w[n]\sim\mathcal{CN}(0,\kappa^{2})$, which is additive white Gaussian noise (AWGN). To establish the optimal performance boundary for parameters estimation, the signal model is first defined. The measured penetration coefficient $y[n]$ can be expressed as
\begin{equation}
y[n] = T(\phi; n) + w[n],
\end{equation}
where $n$ is the index of frequency samples, $\phi = [\varepsilon_r, h, c, d]$ representing the unknown real-valued parameter vector to be estimated, $T(\phi; n)$ denotes the theoretical penetration coefficient calculated from the electromagnetic theoretical model based on the parameter vector to be estimated.

The given expression is the probability density function for a single observation $y[n]$. To correctly represent the joint density for all $N$ sampling points, it should be written as
\begin{equation}
    P(y[1],...,y[n]; \phi) = \prod_{n=1}^{N} \frac{1}{2\pi\kappa^2} \exp\left(-\frac{|y[n] - T(\phi; n)|^2}{2\kappa^2}\right).
\end{equation}

Assume the observed data y[n] follows a probabilistic model with parameters $\phi$. Define the log-likelihood function as
\begin{equation}
    \ln \ell(\phi | y[n]) = -N \ln(2\pi\kappa^2) - \frac{1}{2\kappa^2} \sum |y[n] - T(\phi; n)|^2.
\end{equation}

The calculation of the fisher information matrix (FIM) presupposes the partial derivatives of the log-likelihood function. \(\phi_i\) and \(\phi_j\) denote the \(i\)-th and \(j\)-th elements of the parameter vector \(\phi = [\varepsilon_r, h, c, d]\), respectively, where \(i, j \in \{1, 2, 3, 4\}\). The first-order partial derivative with respect to~$\phi_i$ are expressed as 

    \begin{equation}  % <-- 从 align 改为 equation
    \tag{17}
    \begin{split}     % <-- 在内部嵌套一个 split 环境
        \frac{\partial(\ln \ell)}{\partial\phi_i} &= \frac{1}{\kappa^2} \sum \operatorname{Re}\left[\left(-\frac{\partial T(\phi; n)}{\partial\phi_i}\right)^* (y[n] - T(\phi; n))\right].
    \end{split}       % <-- 结束 split
    \end{equation}    % <-- 结束 equation

 This allows for the derivation of the second-order partial derivative with respect to~$\phi_i$ and~$\phi_j$, as shown in (18). For the specific set of parameters $\phi$ in our model, the FIM can be expressed as

\begin{equation}  % <-- 从 align 改为 equation
\tag{19}
    \begin{split}     % <-- 在内部嵌套一个 split 环境
          \mathrm{I}_{ij} &= -E\left[\frac{\partial^2 (\ln \ell)}{\partial\phi_i \partial\phi_j}\right] \\
    &= \frac{1}{\kappa^2} \sum \operatorname{Re}\left[ \left(\frac{\partial T(\phi; n)}{\partial\phi_i}\right)^* \left(\frac{\partial T(\phi; n)}{\partial\phi_j}\right) \right]. \\
    \end{split}       % <-- 结束 split
    \end{equation}    % <-- 结束 equation

To find the CRLB, we need to compute the inverse of this matrix. The calculation involves several determinants derived from the FIM, which are defined as ${G_{\hat{\varepsilon_r}}}$, ${G_{\hat{h}}}$, ${G_{\hat{c}}}$, ${G_{\hat{d}}}$, and $G$ in (20)-(24).

{
\footnotesize
\begin{equation}
G_{\hat{\varepsilon_r}} = \det\left\{ \sum \mathrm{Re}
\begin{bmatrix}
\dfrac{\partial T_n^*}{\partial \varepsilon_r}\dfrac{\partial T_n}{\partial \varepsilon_r} &
\dfrac{\partial T_n^*}{\partial \varepsilon_r}\dfrac{\partial T_n}{\partial h} &
\dfrac{\partial T_n^*}{\partial \varepsilon_r}\dfrac{\partial T_n}{\partial c} \\[4pt]
\dfrac{\partial T_n^*}{\partial h}\dfrac{\partial T_n}{\partial \varepsilon_r} &
\dfrac{\partial T_n^*}{\partial h}\dfrac{\partial T_n}{\partial h} &
\dfrac{\partial T_n^*}{\partial h}\dfrac{\partial T_n}{\partial c} \\[4pt]
\dfrac{\partial T_n^*}{\partial c}\dfrac{\partial T_n}{\partial \varepsilon_r} &
\dfrac{\partial T_n^*}{\partial c}\dfrac{\partial T_n}{\partial h} &
\dfrac{\partial T_n^*}{\partial c}\dfrac{\partial T_n}{\partial c}

\end{bmatrix}
\right\},
\tag{20}\label{eq:20}
\end{equation}

\begin{equation}
G_{\hat{h}} = \det\left\{ \sum \mathrm{Re}
\begin{bmatrix}
\dfrac{\partial T_n^*}{\partial \varepsilon_r}\dfrac{\partial T_n}{\partial \varepsilon_r} &
\dfrac{\partial T_n^*}{\partial \varepsilon_r}\dfrac{\partial T_n}{\partial c} &
\dfrac{\partial T_n^*}{\partial \varepsilon_r}\dfrac{\partial T_n}{\partial d} \\[4pt]
\dfrac{\partial T_n^*}{\partial c}\dfrac{\partial T_n}{\partial \varepsilon_r} &
\dfrac{\partial T_n^*}{\partial c}\dfrac{\partial T_n}{\partial c} &
\dfrac{\partial T_n^*}{\partial c}\dfrac{\partial T_n}{\partial d} \\[4pt]
\dfrac{\partial T_n^*}{\partial d}\dfrac{\partial T_n}{\partial \varepsilon_r} &
\dfrac{\partial T_n^*}{\partial d}\dfrac{\partial T_n}{\partial c} &
\dfrac{\partial T_n^*}{\partial d}\dfrac{\partial T_n}{\partial d}
\end{bmatrix}
\right\},
\tag{21}\label{eq:21}
\end{equation}

\begin{equation}
G_{\hat{c}} = \det\left\{ \sum \mathrm{Re}
\begin{bmatrix}
\dfrac{\partial T_n^*}{\partial \varepsilon_r}\dfrac{\partial T_n}{\partial \varepsilon_r} &
\dfrac{\partial T_n^*}{\partial \varepsilon_r}\dfrac{\partial T_n}{\partial h} &
\dfrac{\partial T_n^*}{\partial \varepsilon_r}\dfrac{\partial T_n}{\partial d} \\[4pt]
\dfrac{\partial T_n^*}{\partial h}\dfrac{\partial T_n}{\partial \varepsilon_r} &
\dfrac{\partial T_n^*}{\partial h}\dfrac{\partial T_n}{\partial h} &
\dfrac{\partial T_n^*}{\partial h}\dfrac{\partial T_n}{\partial d} \\[4pt]
\dfrac{\partial T_n^*}{\partial d}\dfrac{\partial T_n}{\partial \varepsilon_r} &
\dfrac{\partial T_n^*}{\partial d}\dfrac{\partial T_n}{\partial h} &
\dfrac{\partial T_n^*}{\partial d}\dfrac{\partial T_n}{\partial d}
\end{bmatrix}
\right\},
\tag{22}\label{eq:22}
\end{equation}

\begin{equation}
G_{\hat{d}} = \det\left\{ \sum \mathrm{Re}
\begin{bmatrix}
\dfrac{\partial T_n^*}{\partial \varepsilon_r}\dfrac{\partial T_n}{\partial \varepsilon_r} &
\dfrac{\partial T_n^*}{\partial \varepsilon_r}\dfrac{\partial T_n}{\partial h} &
\dfrac{\partial T_n^*}{\partial \varepsilon_r}\dfrac{\partial T_n}{\partial c} \\[4pt]
\dfrac{\partial T_n^*}{\partial h}\dfrac{\partial T_n}{\partial \varepsilon_r} &
\dfrac{\partial T_n^*}{\partial h}\dfrac{\partial T_n}{\partial h} &
\dfrac{\partial T_n^*}{\partial h}\dfrac{\partial T_n}{\partial c} \\[4pt]
\dfrac{\partial T_n^*}{\partial c}\dfrac{\partial T_n}{\partial \varepsilon_r} &
\dfrac{\partial T_n^*}{\partial c}\dfrac{\partial T_n}{\partial h} &
\dfrac{\partial T_n^*}{\partial c}\dfrac{\partial T_n}{\partial c}
\end{bmatrix}
\right\},
\tag{23}\label{eq:23}
\end{equation}

\begin{equation}
G = \det\left\{ \sum \mathrm{Re}
\begin{bmatrix}
\dfrac{\partial T_n^*}{\partial \varepsilon_r}\dfrac{\partial T_n}{\partial \varepsilon_r} &
\dfrac{\partial T_n^*}{\partial \varepsilon_r}\dfrac{\partial T_n}{\partial h} &
\dfrac{\partial T_n^*}{\partial \varepsilon_r}\dfrac{\partial T_n}{\partial c} &
\dfrac{\partial T_n^*}{\partial \varepsilon_r}\dfrac{\partial T_n}{\partial d} \\[4pt]
\dfrac{\partial T_n^*}{\partial h}\dfrac{\partial T_n}{\partial \varepsilon_r} &
\dfrac{\partial T_n^*}{\partial h}\dfrac{\partial T_n}{\partial h} &
\dfrac{\partial T_n^*}{\partial h}\dfrac{\partial T_n}{\partial c} &
\dfrac{\partial T_n^*}{\partial h}\dfrac{\partial T_n}{\partial d} \\[4pt]
\dfrac{\partial T_n^*}{\partial c}\dfrac{\partial T_n}{\partial \varepsilon_r} &
\dfrac{\partial T_n^*}{\partial c}\dfrac{\partial T_n}{\partial h} &
\dfrac{\partial T_n^*}{\partial c}\dfrac{\partial T_n}{\partial c} &
\dfrac{\partial T_n^*}{\partial c}\dfrac{\partial T_n}{\partial d} \\[4pt]
\dfrac{\partial T_n^*}{\partial d}\dfrac{\partial T_n}{\partial \varepsilon_r} &
\dfrac{\partial T_n^*}{\partial d}\dfrac{\partial T_n}{\partial h} &
\dfrac{\partial T_n^*}{\partial d}\dfrac{\partial T_n}{\partial c} &
\dfrac{\partial T_n^*}{\partial d}\dfrac{\partial T_n}{\partial d}
\end{bmatrix}
\right\}.
\tag{24}\label{eq:24}
\end{equation}
}

The estimated CRLB has express as

\begin{equation} 
\tag{25}
\left\{ 
\begin{alignedat}{2} 
    &\operatorname{CRLB}(\hat{\varepsilon}_r) &&= [I^{-1}(\varepsilon_r, h, c, d)]_{11} = \frac{G_{\hat{\varepsilon}_r}}{\kappa^2 G},\\ 
    &\operatorname{CRLB}(\hat{h})           &&= [I^{-1}(\varepsilon_r, h, c, d)]_{22} = \frac{G_{\hat{h}}}{\kappa^2 G}, \\ 
    &\operatorname{CRLB}(\hat{c})           &&= [I^{-1}(\varepsilon_r, h, c, d)]_{33} = \frac{G_{\hat{c}}}{\kappa^2 G}, \\ 
    &\operatorname{CRLB}(\hat{d})           &&= [I^{-1}(\varepsilon_r, h, c, d)]_{44} = \frac{G_{\hat{d}}}{\kappa^2 G}. 
\end{alignedat} 
\right. 
\end{equation}

Due to the nonlinear nature of the model and the complexity of the parameter space, the second-order partial derivatives of the log-likelihood function with respect to the parameters are difficult to solve analytically. Therefore, this paper employs the nine-point central difference scheme (26) to numerically calculate the second-order partial derivatives.

In this paper, the inverse of the FIM is computed using a numerical matrix inversion method to obtain the CRLB for the parameter estimates. The resulting CRLB serves as a theoretical lower bound on performance and is used to evaluate the estimation accuracy of the proposed FA-PSO algorithm.

\section{Results and Discussion}

In this section, we present and analyze the experimental results to provide a comprehensive evaluation of the free-space measurement method and the FA-PSO algorithm for extracting complex permittivity and material thickness. 

%Subsection A focuses on the validation of the method using standard building materials, Subsection B examines the accuracy of thickness measurement. Subsection C investigates the influence of material thickness on inversion reliability. Subsection D compares the estimation errors of the proposed algorithm with the CRLB in thickness-varying simulations.

\subsection{Complex Permittivity Extraction from Building Materials}

% --- 请确保在导言区（preamble）加载了这两个宏包 ---
% \usepackage{booktabs}
% \usepackage{siunitx}
% ----------------------------------------------------

\begin{figure}[!t]
\centerline{\includegraphics[width=0.95\columnwidth]{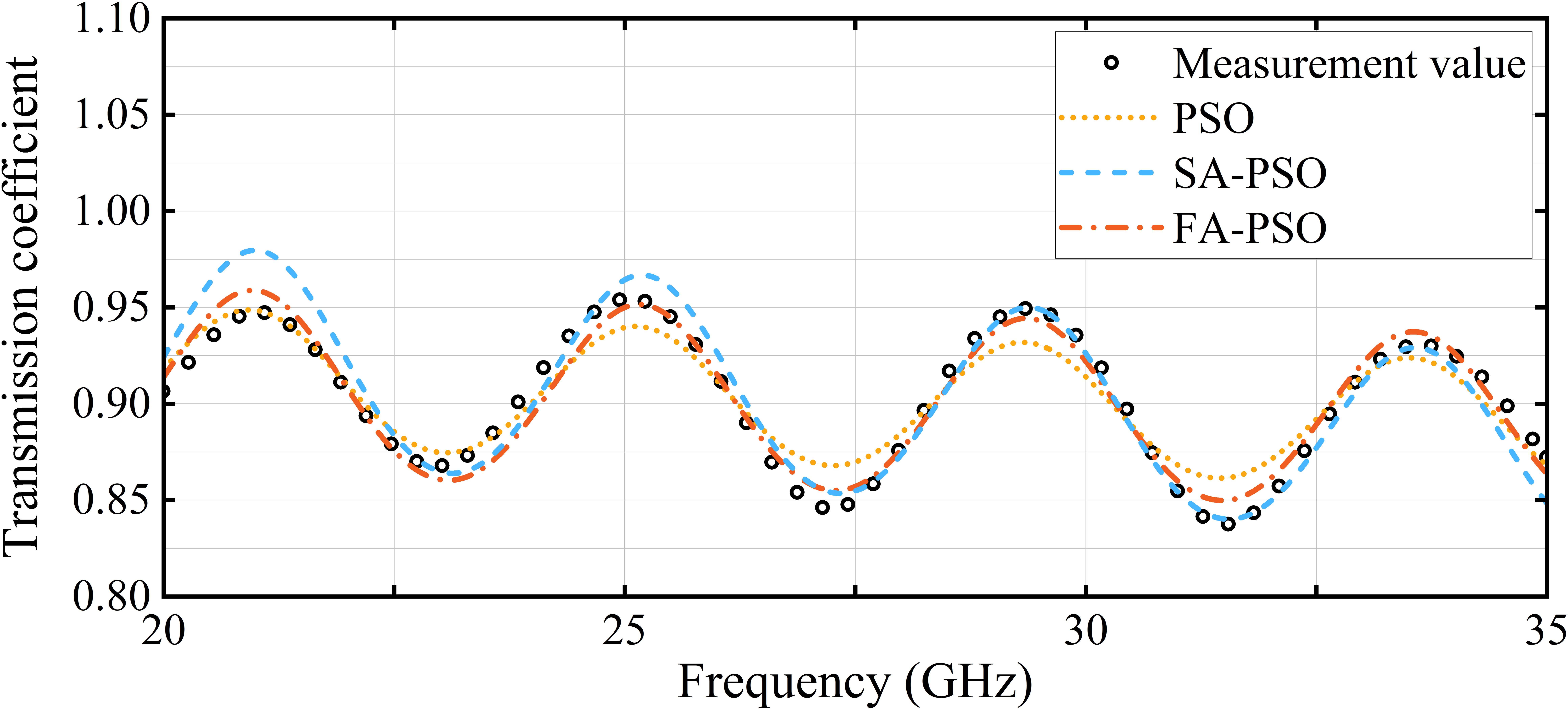}}
\caption{Penetration coefficient from PET measurement at normal incidence.}
\label{fig4}
\end{figure}

\begin{table}[!t] 
\caption{Permittivity of PET measured by different methods.} 
\label{tab:pet_permittivity} 
\centering % 确保整个表格环境在页面上居中

% --- 通过下面的命令来调整列间距 ---
% 您可以修改 12pt 这个值来获得想要的间距
\setlength{\tabcolsep}{12pt} 

% --- 列定义更新为 c c c，所有列都将居中对齐 ---
\begin{tabular}{ccc} 
\toprule
Sources   & Permittivity & Method \\ 
\midrule
$[31]$ & 2.672        & Waveguide method \\ 
$[32]$ & 2.500        & Waveguide method \\ 
$[33]$  & 2.884        & Interferometry method \\ 
 $[34]$ & 3.082        & Resonant cavity method \\ 
This work & 2.878        & Free-space method \\ 
\bottomrule
\end{tabular} 
\end{table}

\begin{figure}[!t]
\centering
    \begin{minipage}{1\linewidth}
        \centering
        % 第一行图片
        \begin{subfigure}{\textwidth}
            \centering
            \includegraphics[width=1\textwidth]{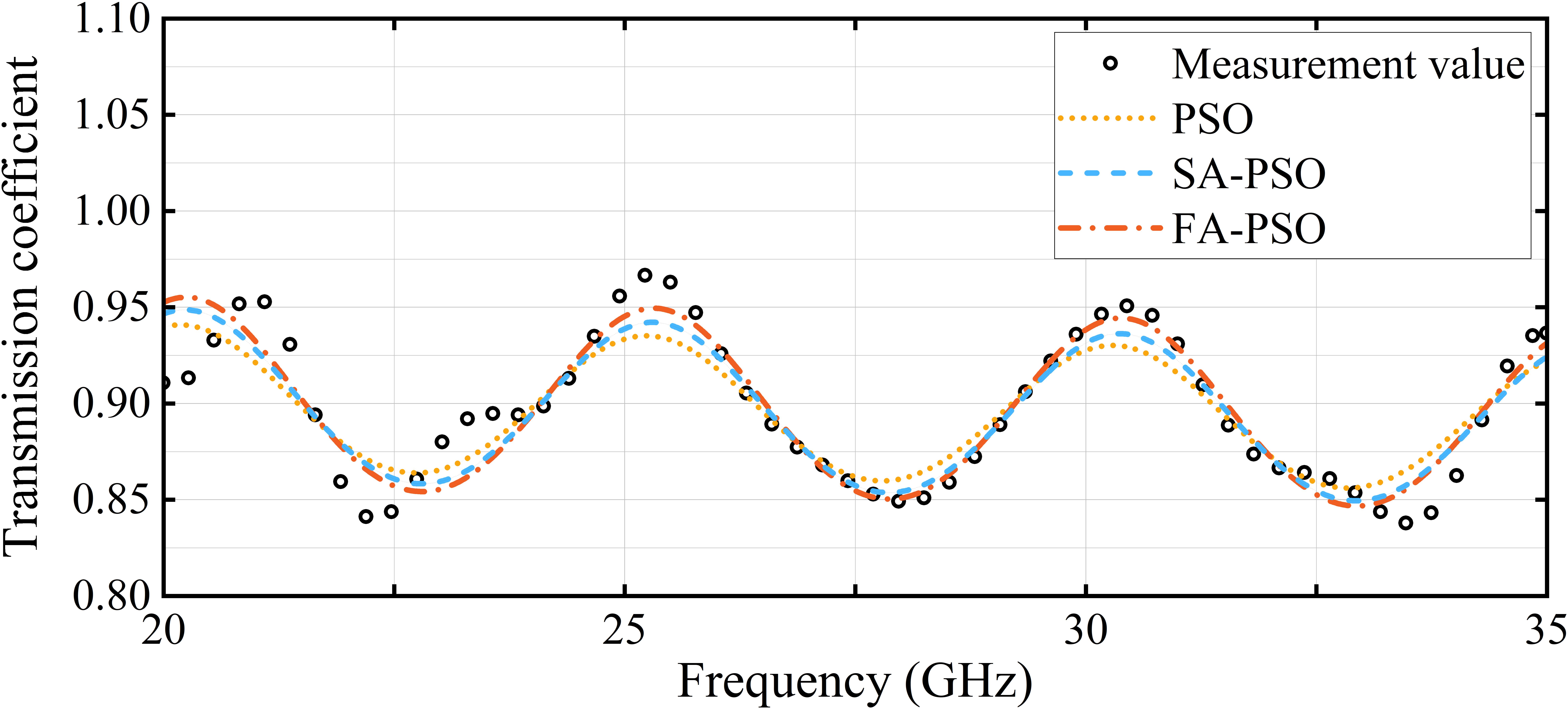}
            \caption{}
            %\label{fig:sub1}
        \end{subfigure}
        
        % 第二行图片
        \begin{subfigure}{\textwidth}
            \centering
            \includegraphics[width=1\textwidth]{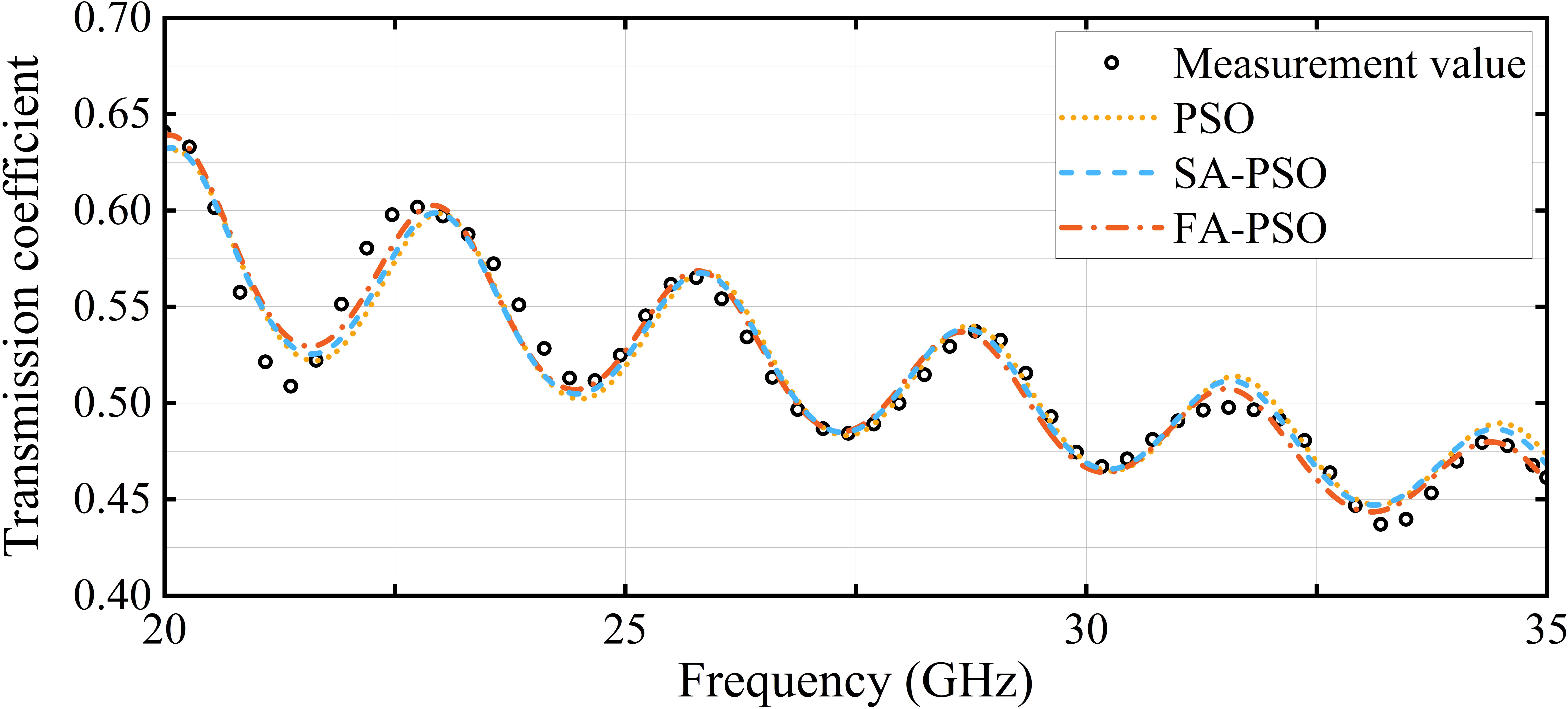}
            \caption{}
            %\label{fig:sub2}
        \end{subfigure}
        
        % 第三行图片
        \begin{subfigure}{\textwidth}
            \centering
            \includegraphics[width=1\textwidth]{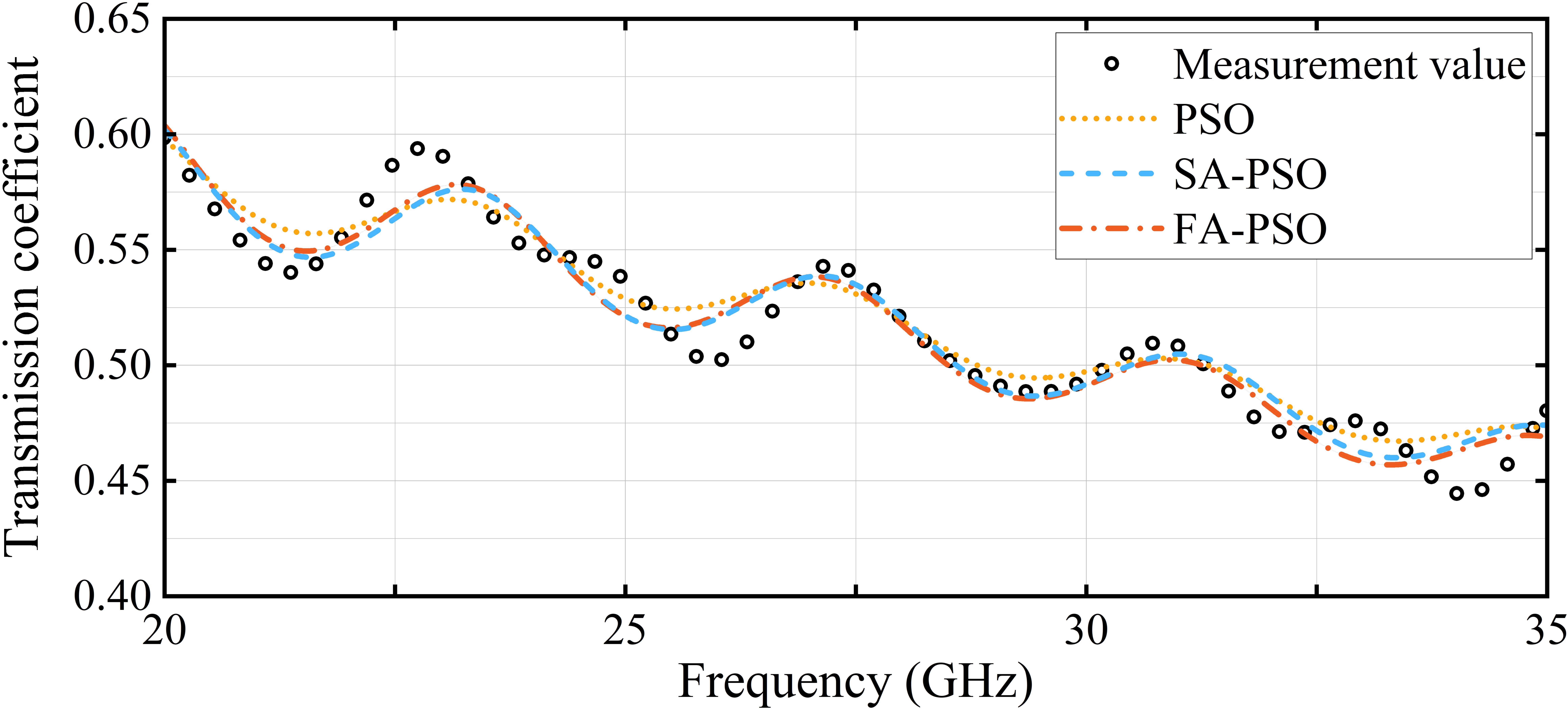}
            \caption{}
            %\label{fig:sub3}
        \end{subfigure}
    \end{minipage}
    \hfill
\caption{Measured and calculated penetration  under incident angle of 0°. (a) Acrylic. (b) Rubber. (c) Bakelite.}
\label{fig7}
\end{figure}

\begin{table}[!t] 
\caption{Electromagnetic Parameters and Error Analysis.} 
\label{tab:em_params} 
\centering 

% 调整行高，让内容更舒展
\renewcommand{\arraystretch}{1.3} 

% 定义5个居中列
% 我们将最后一列也用 S 列处理，以更好地显示科学计数法
\begin{tabular}{c S[table-format=1.3] S[table-format=1.4] S[table-format=1.3] S[table-format=1.2e-2]} 
\toprule
\multirow{2}{*}{Building materials} & {\multirow{2}{*}{$\varepsilon'$}} & \multicolumn{2}{c}{$\sigma=cf^d (\mathrm{S/m})$} & {\multirow{2}{*}{RMSE}} \\ 
\cmidrule(lr){3-4} % <--- 使用 \cmidrule 替代 \cline，并左右留出空隙
 & & {c} & {d} & \\ 
\midrule
Acrylic & 2.700 & 0.0134 & 0.551 & 1.38e-2 \\ 
Rubber & 5.683 & 0.1010 & 1.0017 & 9.44e-3 \\ 
Bakelite boards & 3.796 & 0.124 & 0.839 & 1.02e-2 \\ 
\bottomrule
\end{tabular} 
\end{table}

\begin{figure}[!t]
\centering
    \begin{minipage}{1\linewidth}
        \centering
        % 第一行图片
        \begin{subfigure}{\textwidth}
            \centering
            \includegraphics[width=1\textwidth]{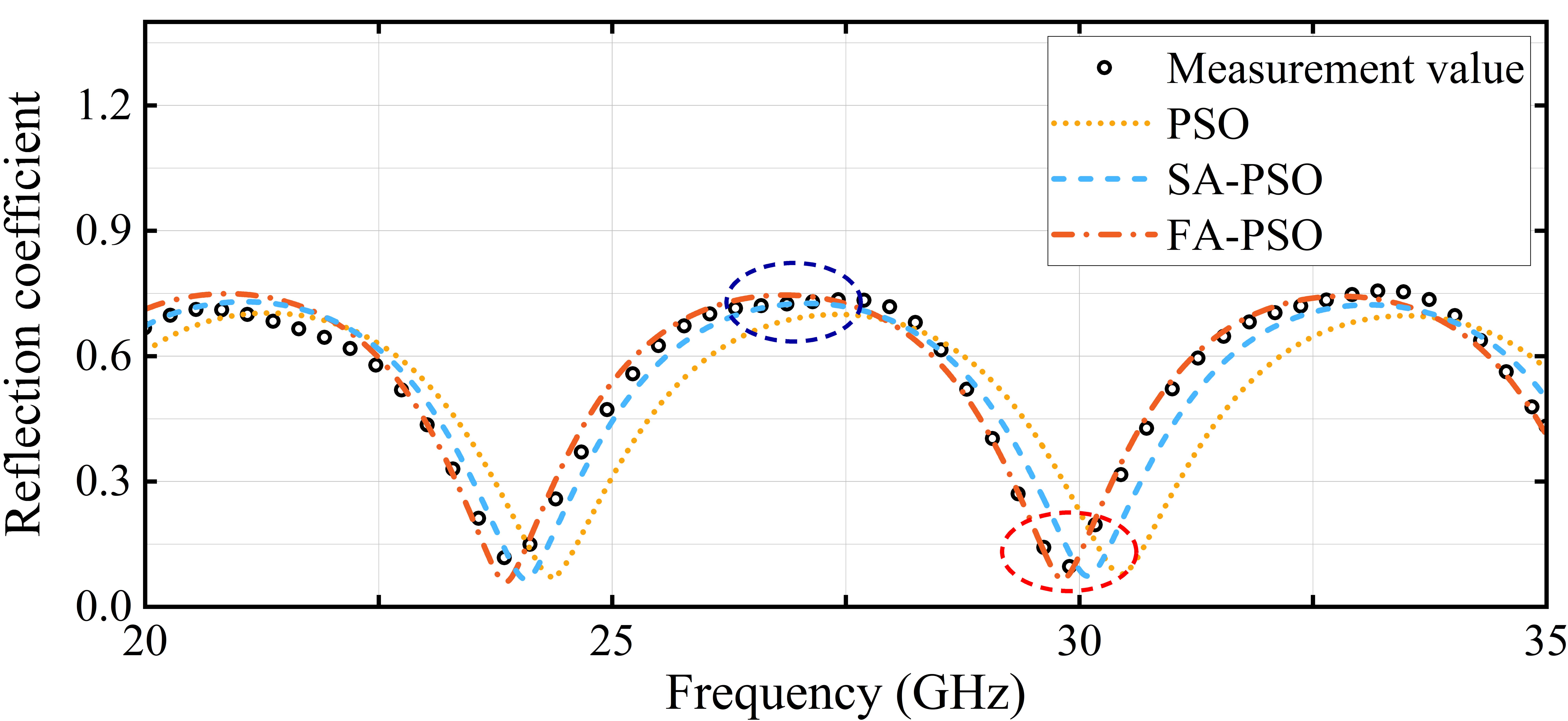}
            \caption{}
            %\label{fig:sub1}
        \end{subfigure}
        
        % 第二行图片
        \begin{subfigure}{\textwidth}
            \centering
            \includegraphics[width=1\textwidth]{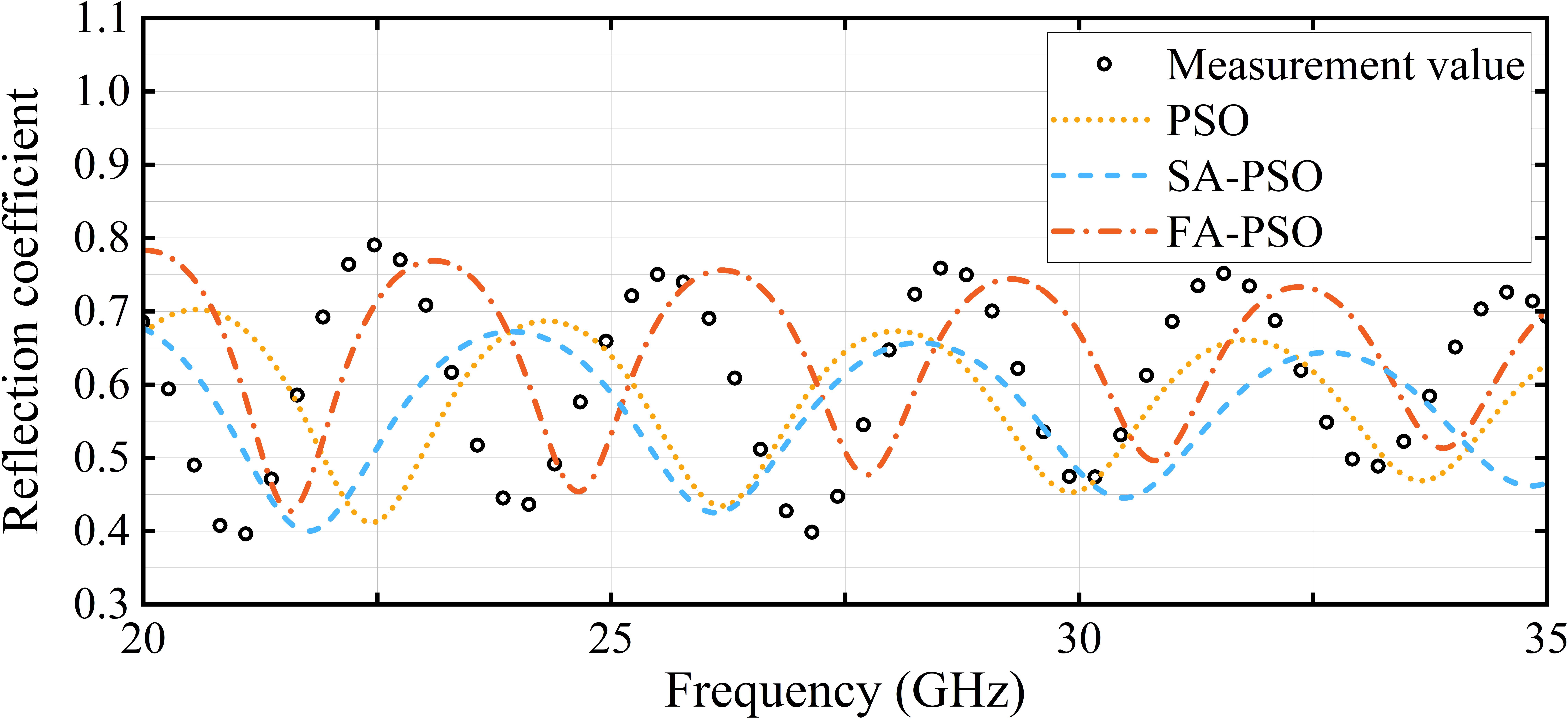}
            \caption{}
            %\label{fig:sub2}
        \end{subfigure}
        
        % 第三行图片
        \begin{subfigure}{\textwidth}
            \centering
            \includegraphics[width=1\textwidth]{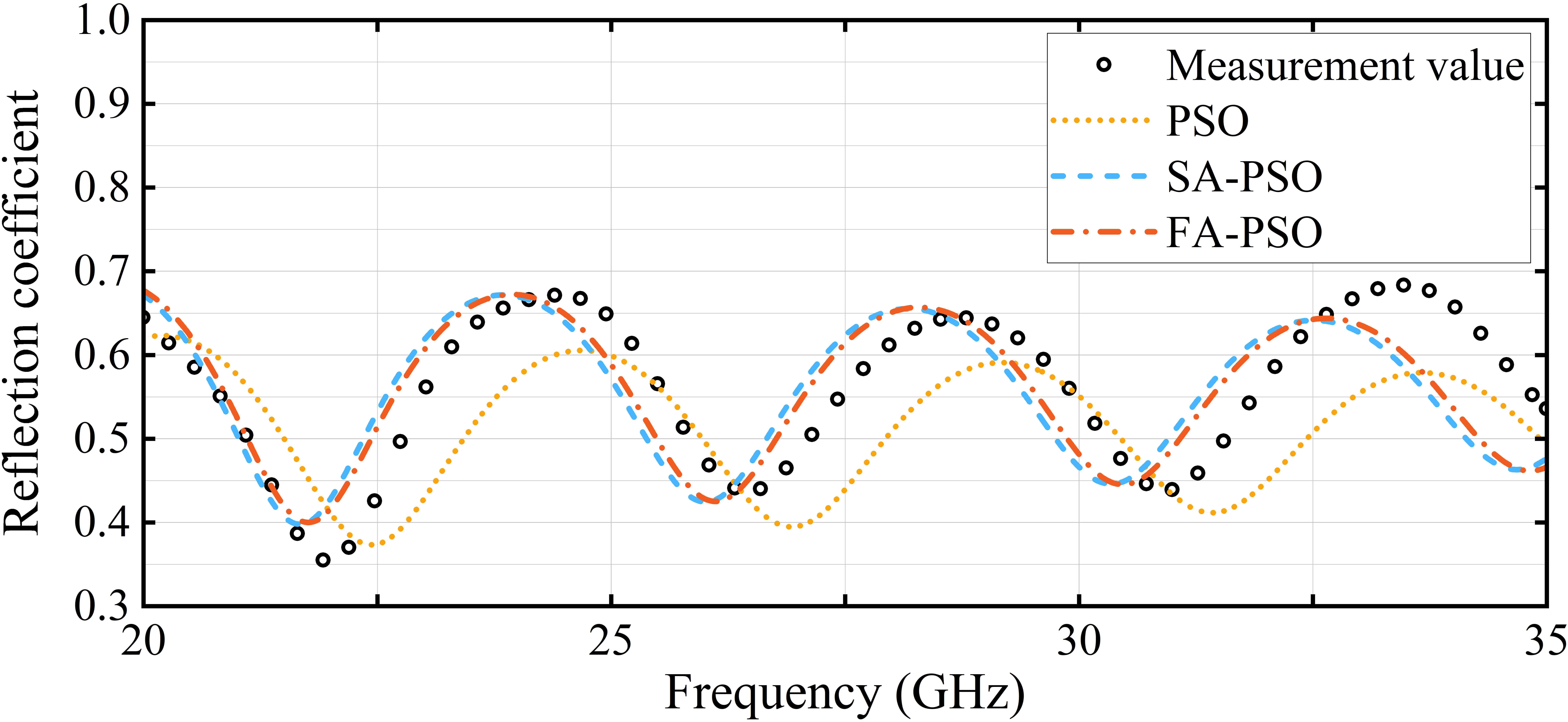}
            \caption{}
            %\label{fig:sub3}
        \end{subfigure}
    \end{minipage}
    \hfill
\caption{Measured and calculated reflection coefficient under incident angle of 60°. (a) Acrylic. (b) Rubber. (c) Bakelite.}
\label{fig8}
\end{figure}

To validate the effectiveness of the proposed measurement procedure and extraction method, a uniform PET plate with dimensions of 30 cm in length, 30 cm in width, and 2 cm in thickness was first selected as the standard reference material. As a homogeneous isotropic material with well-characterized electromagnetic properties, PET has a thoroughly validated permittivity, making it a reliable reference for the FA-PSO algorithm's measurement results. The extracted results and data from other sources, such as the waveguide method, interferometry method, resonant cavity method, and free-space method, are shown in Table \Rmnum{1} and Fig.~\ref{fig4}. The comparison between our extracted results and the data from other methods indicates similar numerical values, thereby validating the accuracy of the measurement and inversion process. In addition, Fig.~\ref{fig4} also presents the inversion results obtained using PSO and simulated annealing and particle swarm optimization (SA-PSO). After confirming the reliability of the measurement and extraction results, we used the FA-PSO algorithm to extract the permittivity, conductivity, and thickness of the material.

Fig. \ref{fig7} presents the curves of the penetration coefficient amplitude versus frequency for the above three materials. For comparison, the corresponding inversion curves obtained using the PSO, the FA-PSO, and the SA-PSO algorithms are also displayed. The results indicate that all samples exhibit obvious frequency characteristics. The differences in response between different materials are significant, which directly reflects their different internal structures. Specifically, compared to bakelite with more complex internal structures, acrylic and rubber materials with relatively simple structures exhibit smoother changes in penetration coefficients with frequency. Subsequently, the measured penetration data were inverted using the FA-PSO algorithm to obtain the complex dielectric constants of the three materials. The material parameters obtained from the inversion, including dielectric constants, electrical conductivities, estimated thicknesses, and RMSE values for each material sample, are summarized in Table~\Rmnum{2} using this method. The low RMSE values indicate that the proposed FA-PSO inversion model is highly consistent with the experimental data, thereby verifying the accuracy and reliability of this method in simultaneously evaluating the dielectric properties and geometric dimensions of materials.

To verify the validity of the complex permittivity extracted by FA-PSO, we also measured the reflection coefficient at an incident angle of 60°. The permittivity, conductivity, and thickness extracted by the FA-PSO algorithm theoretical reflection coefficient, which was then compared with the reflection coefficient calculated directly from the measured data. As shown in Fig. \ref{fig8}, the trends in the reflection coefficients of the various materials match well, verifying the reliability of the algorithm. Notably, the reflection coefficient curve of the uniform acrylic slab exhibits a series of periodic deep fades, which even approach zero at specific resonant frequencies. This phenomenon is attributed to destructive interference induced by the multi-path effect of electromagnetic waves between the front and rear surfaces of the dielectric slab. The spacing of these resonant frequencies is directly related to the dielectric constant and physical thickness of the material. By performing an inverse fast Fourier transform (IFFT) on the 3 dB bandwidth around the channel impulse response at 27.5 GHz and 29.6 GHz, with a bandwidth of 0.5 GHz, the corresponding impulse responses are obtained, as shown in Fig. 10 (a). The impulse responses indicate that when the reflection coefficient approaches zero, the peak of the impulse response also decreases and becomes close to the noise floor. The horizontal axis in Fig. 11 is the relative delay of the estimated impulse response, not absolute time.

\begin{figure}[tbp]
\centering
\begin{subfigure}[b]{0.24\textwidth}\centering
  \includegraphics[width=\linewidth]{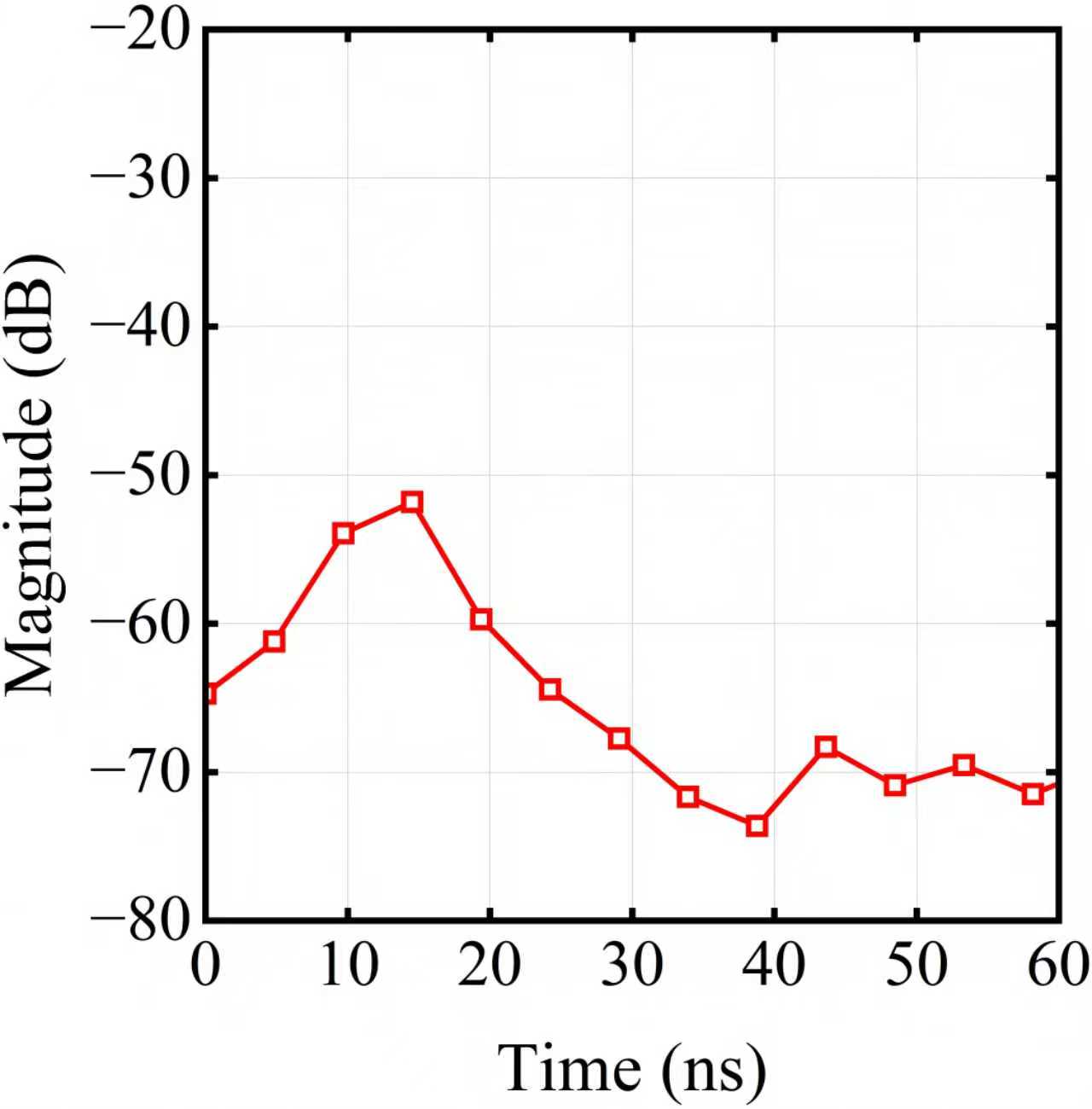}
  \subcaption{}\label{fig:sub_a}
\end{subfigure}\hfill
\begin{subfigure}[b]{0.24\textwidth}\centering
  \includegraphics[width=\linewidth]{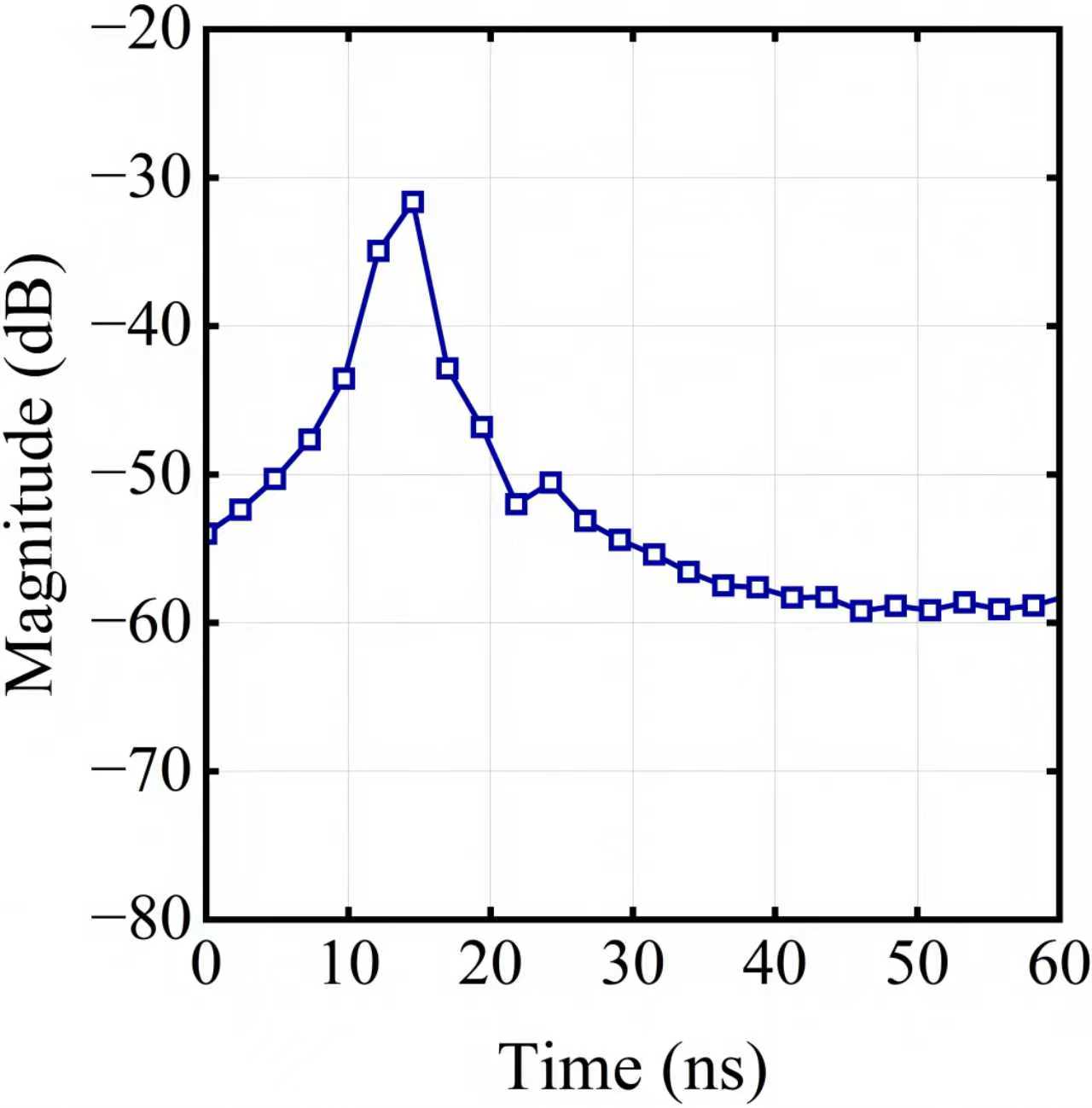}
  \subcaption{}\label{fig:sub_b}
\end{subfigure}
\caption{Impulse responses at  channel impulse response at~27.5 GHz and 29.6 GHz in Fig. 10 (a). (a) 27.5 GHz. (b) 29.6 GHz}
\label{fig11}
\end{figure}

\subsection{FA-PSO Algorithm for Building Material Thickness Measurement}

\begin{table}[!t]
\caption{Measurement and Calculation Thickness of MUT.}
\label{table3}
\centering
\renewcommand{\arraystretch}{1.2}
\setlength{\tabcolsep}{5pt}

\begin{tabular}{cccc}
\toprule
MUT (mm) & Calculated thickness & Measured thickness & Error \\
\midrule
Acrylic  & 18.00 & 18.39 & 0.39 \\
Rubber   & 21.90 & 21.38 & 0.52 \\
Bakelite & 19.80 & 20.12 & 0.32 \\
\bottomrule
\end{tabular}
\end{table}

Our previous research has shown that the thickness of the material has a critical impact on spatially averaged capacity. However, in the measurement scenarios we anticipate for future wireless performance of buildings, in order to achieve rapid assessment, it is necessary to avoid the complex process of measuring the thickness of the building structure. It should be noted that the electromagnetic parameter extraction algorithm used in this study also has the function of accurately measuring the thickness of the material, which lays the technical foundation for future rapid assessment platforms for wireless performance in buildings. To verify the accuracy of the algorithm's thickness measurements, we compared the thickness calculation results with the actual measured values, which were obtained directly using a vernier caliper in Fig. \ref{fig9}. The inversion results and actual measured values are summarized in Table \ref{table3}. As can be seen, the errors between the inversion and actual measured results are both less than 0.6 mm, demonstrating the high accuracy of this algorithm in thickness measurements.

\subsection{FA-PSO Based on Gaussian Initial Particle Distribution}

The standard PSO algorithm typically employs a uniform distribution for population initialisation. However, for highly non-linear least-squares problems such as electromagnetic parameter inversion, the objective function often contains numerous local minima. The distribution of the initial population within the parameter space decisively influences the algorithm's convergence speed and solution accuracy. The algorithm's performance is critically dependent on its initial conditions. An initial particle swarm distant from the global optimum is highly susceptible to stagnation or premature convergence. While an exhaustive search of the parameter space guarantees finding the global optimum, its prohibitive computational cost renders it impractical. Therefore, developing an efficient initialization strategy is crucial. This strategy directs new particle groups to cluster in high-probability regions, thereby focusing computational resources on the areas most likely to contain the optimal solution.

\begin{figure}[!t]
\centerline{\includegraphics[width=0.9\columnwidth]{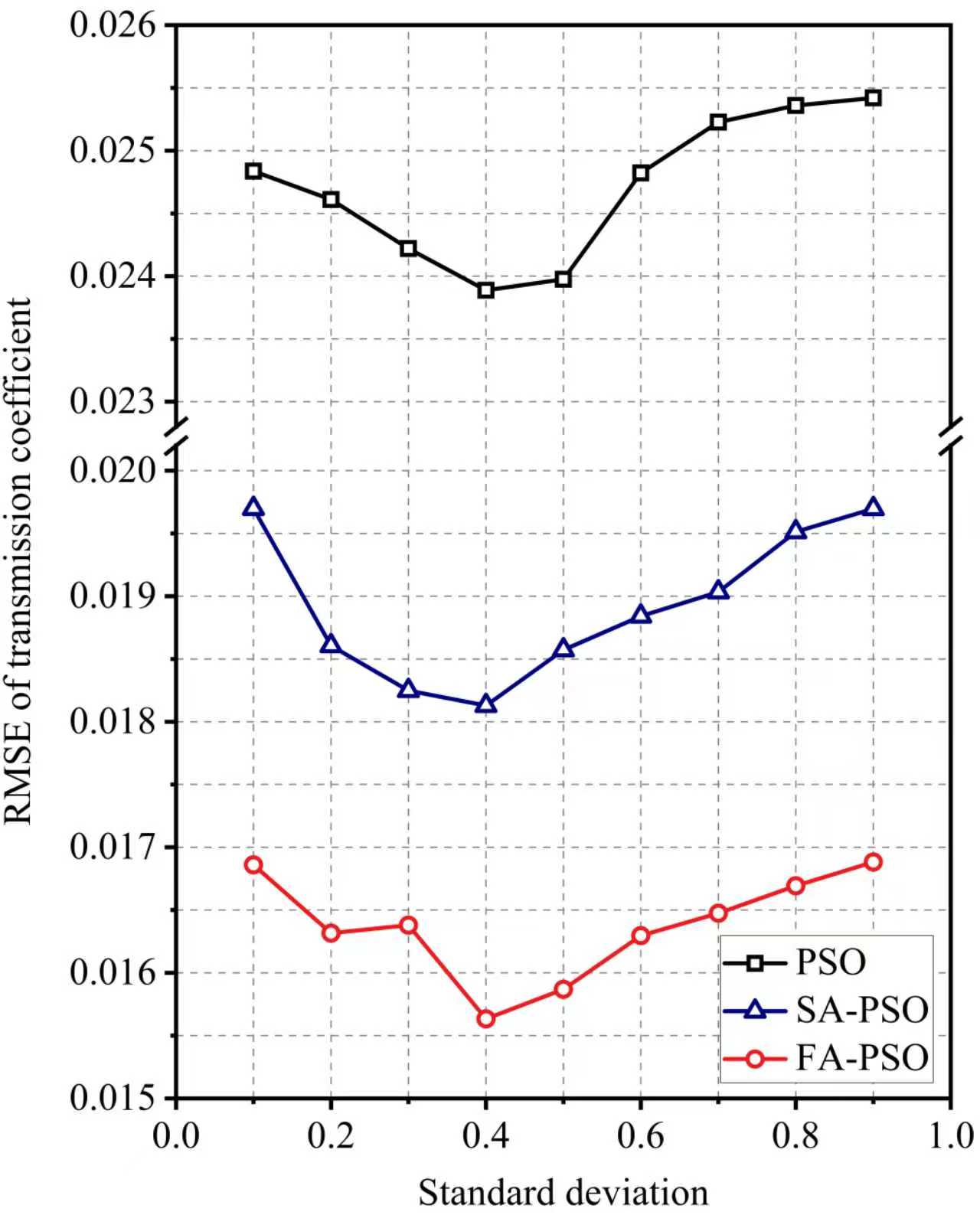}}
\caption{Variation of RMSE with input standard deviation for the PSO, SA-PSO, and FA-PSO algorithms. }
  \label{fig10}
\end{figure}

% --- 将这个代码块用于您的 2x2 图片排列 ---
\begin{figure*}[tbp] % 推荐使用 [tbp] (top, bottom, page)
    \centering % 整体居中

    % --- 第一行图片 ---
    \begin{subfigure}[b]{0.45\textwidth}
        \centering
        \includegraphics[width=\linewidth]{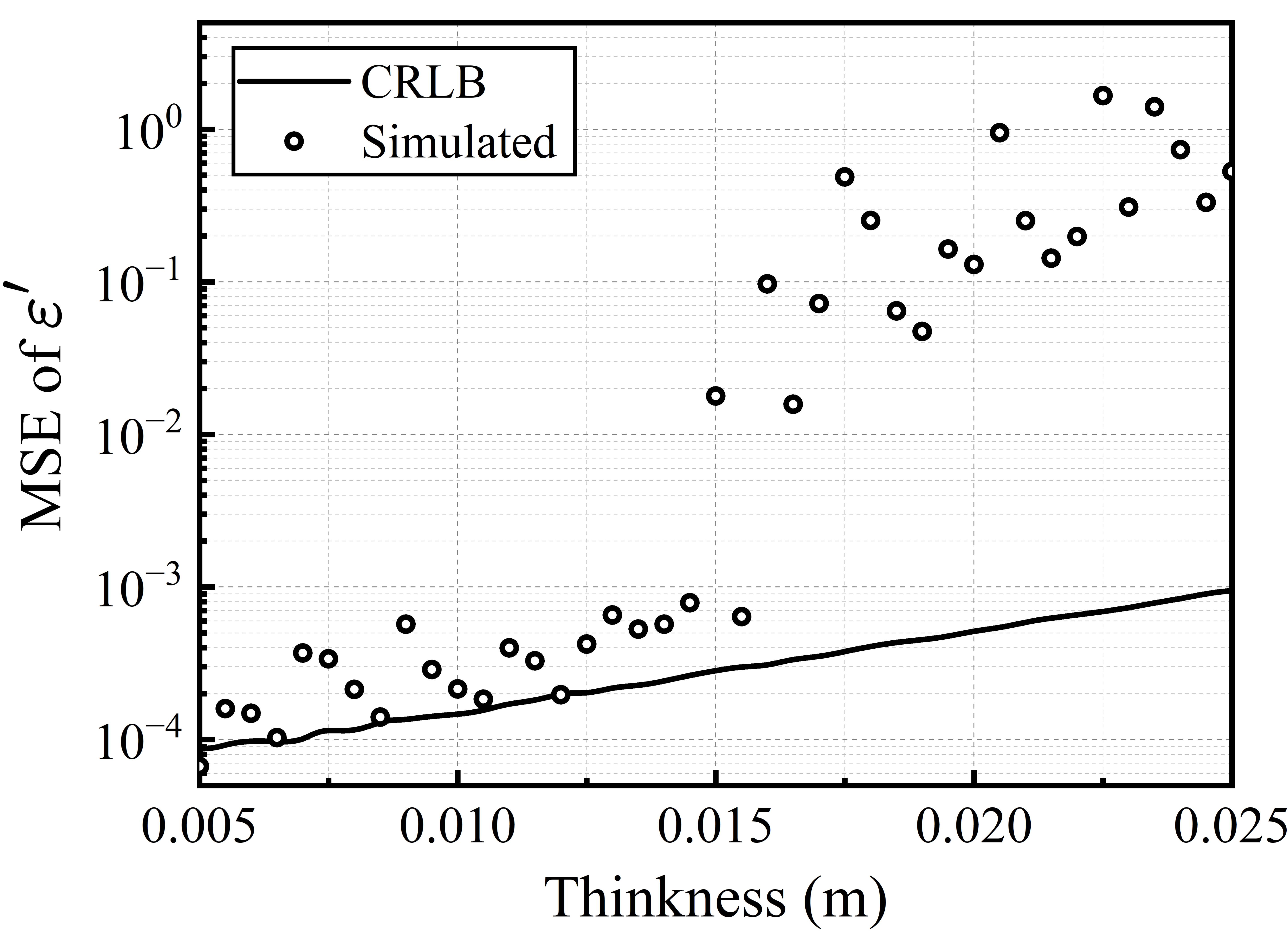} % <-- 替换为您的第一张图片路径
        \subcaption{} % <-- 图片a的子标题
        \label{fig:sub_a}
    \end{subfigure}
    \hspace{1em} % 在两张图片之间创建弹性水平间距
    \begin{subfigure}[b]{0.45\textwidth}
        \centering
        \includegraphics[width=\linewidth]{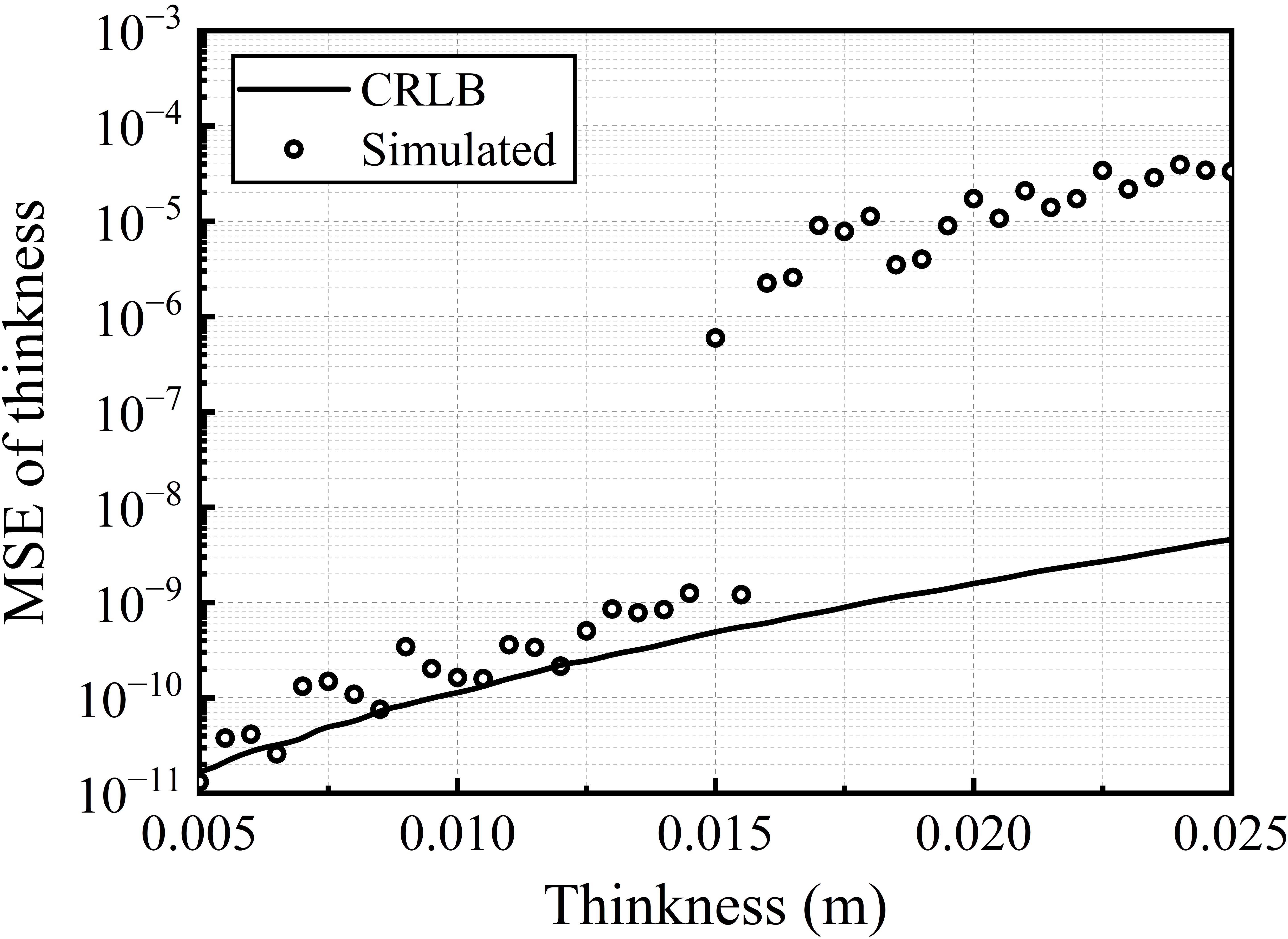} % <-- 替换为您的第二张图片路径
        \subcaption{} % <-- 图片b的子标题
        \label{fig:sub_b}
    \end{subfigure}

    % 在两行图片之间增加一些垂直间距
   % \vspace{1em} 

    % --- 第二行图片 ---
    \begin{subfigure}[b]{0.45\textwidth}
        \centering
        \includegraphics[width=\linewidth]{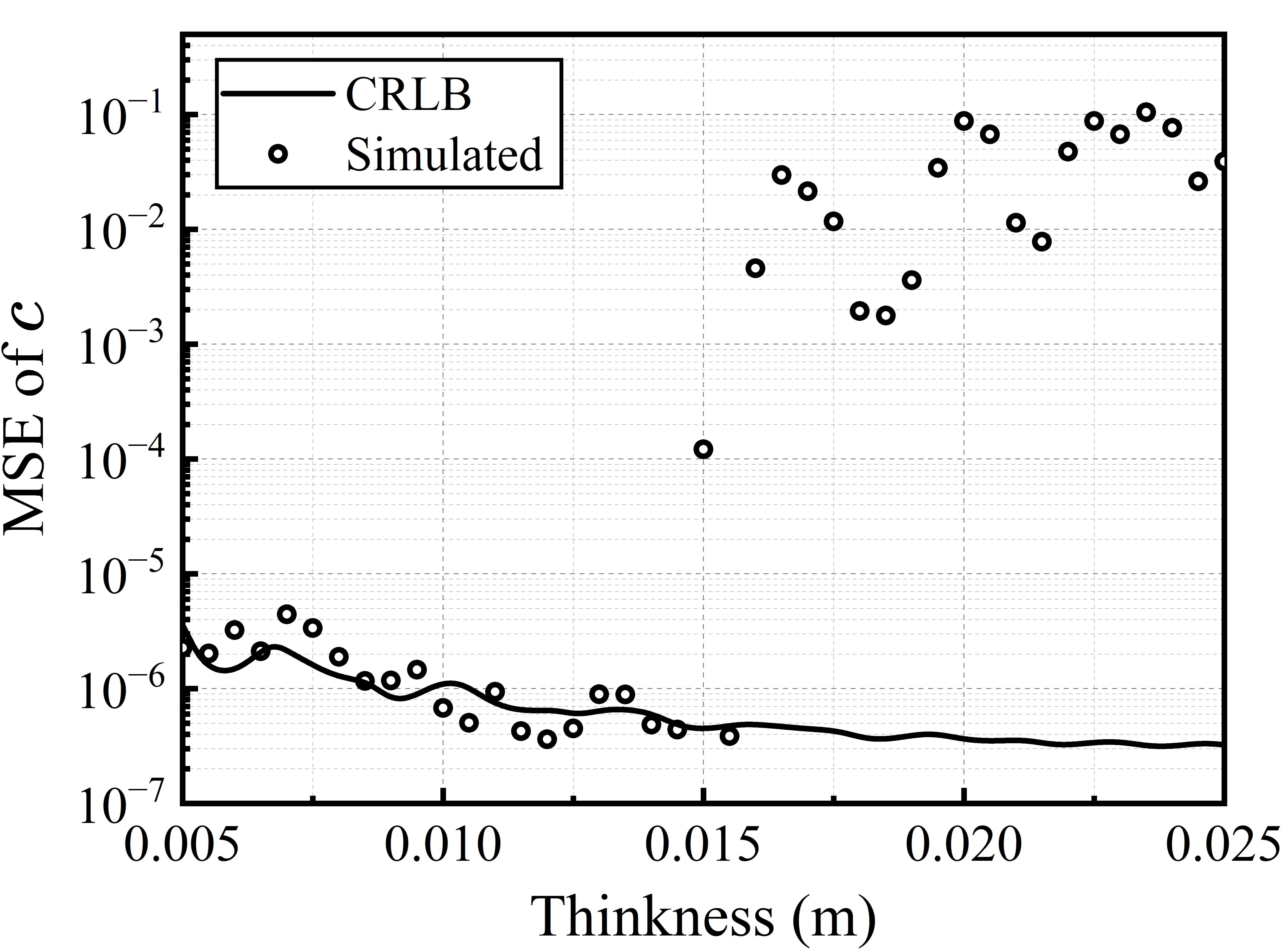} % <-- 替换为您的第三张图片路径
        \subcaption{} % <-- 图片c的子标题
        \label{fig:sub_c}
    \end{subfigure}
    \hspace{1em} % 在两张图片之间创建弹性水平间距
    \begin{subfigure}[b]{0.45\textwidth}
        \centering
        \includegraphics[width=\linewidth]{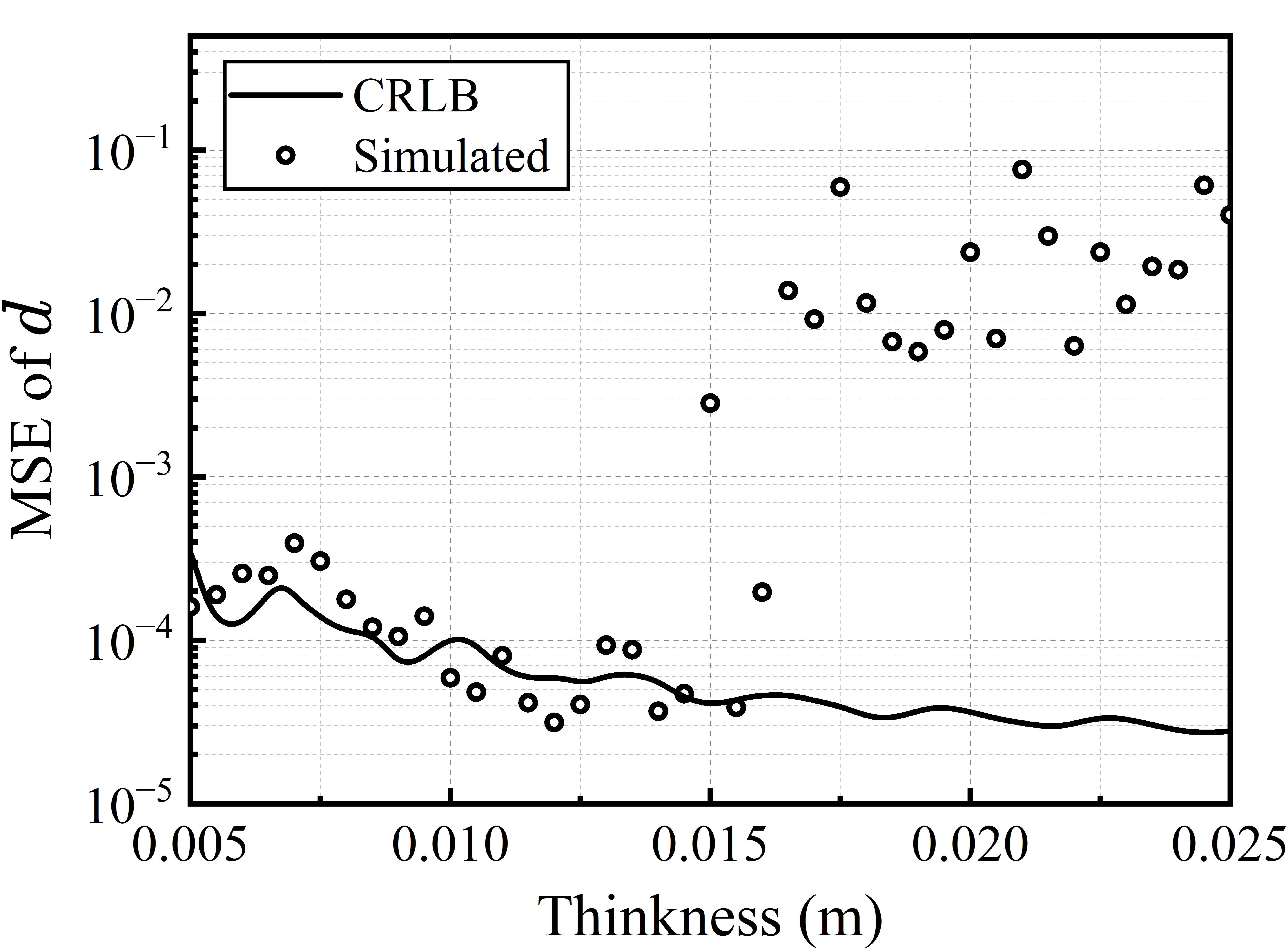} % <-- 替换为您的第四张图片路径
        \subcaption{} % <-- 图片d的子标题
        \label{fig:sub_d}
    \end{subfigure}

    % --- 整个图的总标题和总标签 ---
    \caption{Variation of MSE with MUT Thickness. (a) Permittivity. (b) Thickness. (c) \textit{c} in (8). (d) \textit{d} in (8). }
    \label{fig11}
\end{figure*}

To reduce computational complexity, a reasonable prior estimate for the dielectric constant is established based on typical parameter ranges for building materials. Based on the prior empirical value of approximately 5.5 for the relative dielectric constant of rubber, the mean $\mu$ of the Gaussian distribution is fixed. The RMSE is employed as the evaluation metric to investigate the impact of variations in the standard deviation $\tau$ on model performance. Concurrently, to evaluate algorithmic performance, PSO,  SA-PSO, and FA-PSO are selected as comparative benchmarks. In the experimental design, all algorithms employ a population size of 1000 and a maximum iteration count of 100. To mitigate randomness effects, each algorithm is executed 100 times, with the convergence value derived from the average of these results.

The experimental results are shown in Fig. \ref{fig10}, which compares the absolute errors of the three algorithms under different noise standard deviations. Analysis reveals that the fundamental PSO algorithm exhibits relatively high error due to its inherent tendency to become trapped in local optima, while SA-PSO demonstrates comparable error owing to the random nature of its global search. In contrast, the FA-PSO algorithm, which integrates the strengths of both FA and PSO, significantly reduces absolute error through optimised search strategies. Concurrently, experimental results indicate that as the $\tau$ value increases, the RMSE of the inversion results initially decreases and then increases, reaching a minimum when $\tau$ = 0.4. Therefore, a value of $\tau$ = 0.4 was ultimately selected as the hyperparameter for the prior Gaussian distribution of the dielectric constant. This strategy effectively guides the particle swarm algorithm to search high-potential energy regions through empirical knowledge, thereby enhancing its convergence efficiency and global optimisation capability.

\subsection{Investigating the Influence of Material Thickness on Inversion Reliability}
In the simulation experiment section, we employ the CRLB derived in the preceding section as the theoretical benchmark for evaluating the performance of the proposed algorithm. The validity of the algorithm is verified by comparing the closeness of the estimation error to the CRLB. Rubber was selected as the study subject. Using the measured electromagnetic parameters of rubber obtained in Section IV, we simulated the penetration coefficient of rubber plates at varying thicknesses ranging from 0.5 cm to 2.5 cm (with 0.5 mm intervals) across the 20-35 GHz frequency band. We generated simulated measurement data by adding complex Gaussian white noise with a standard deviation of 0.07 to the theoretical penetration coefficients. Subsequently, the proposed Gaussian initialisation FA-PSO algorithm was employed to perform parameter inversion on the simulated data.

From Fig. \ref{fig11}, it can be clearly observed that within the thinner thickness range (0.5 cm to 1.5 cm), the MSE for all parameters closely follows its corresponding CRLB curve. This tight fit between RMSE and CRLB indicates that under these conditions, the proposed FA-PSO algorithm is statistically efficient, meaning its estimation performance approaches theoretical optimality. It is worth noting that estimation errors slightly widen and deviate from the CRLB as thickness increases. We hypothesise this arises because increased thickness intensifies multiple reflections of electromagnetic waves within the material, thereby amplifying the nonlinear coupling and interdependence among parameters within the model. This heightened complexity reduces the objective function's sensitivity to certain parameters-a phenomenon commonly observed in iterative or heuristic algorithms. Particularly under conditions of reduced signal-to-noise ratio and heightened parameter interference, the estimation error exhibits diminished convergence towards theoretical limits, reflecting an unavoidable engineering trade-off. Nevertheless, the deviation remains within acceptable engineering parameters, demonstrating the algorithm's robustness and reliability in practical applications.

The above results clearly indicate that the proposed FA-PSO algorithm demonstrates excellent performance when processing plate-shaped materials with thicknesses less than 1.5 cm, whereas its measurement accuracy deteriorates significantly for thicker MUTs. This finding carries important implications, revealing that current electromagnetic parameter measurement techniques based on the free-space method are only applicable to materials of limited thickness and still exhibit limitations in accurately characterizing thicker structures such as walls. Therefore, future research should focus on continuously improving the robustness and performance of inversion algorithms to raise the effective measurement measurement threshold. Thereby expanding the application boundaries of the free-space method.

\section{Conclusions}

%This study developed a FA-PSO algorithm for electromagnetic parameter extraction. It overcomes the key limitations of local optima and premature convergence in traditional inversion methods. The proposed algorithm accurately extracted the penetration coefficients of three typical building materials, including acrylic, rubber, and bakelite, in the 20–35 GHz frequency band. The extraction achieved a low RMSE of $9.44 \times 10^{-3}$. Additionally, the technique can simultaneously and accurately estimate material thickness, with an error of less than 0.6 mm compared to the measured values. Furthermore, to enhance algorithmic performance, the initial particle swarm of the FA-PSO algorithm was optimised. Simulation results indicate that the inversion accuracy approaches the CRLB. The proposed approach will facilitate building material selection in terms of wireless performance for civil engineer.

In this paper, we developed a FA-PSO algorithm to extraction the permittivity, conductivity, and thickness of building materials using the free-space method. The FA-PSO algorithm accurately extracted the permittivity, conductivity, and thickness of building materials from the penetration coefficients of three typical building materials, including acrylic, rubber, and bakelite, in the 20–35 GHz frequency band. The extraction achieved a low RMSE of $9.44 \times 10^{-3}$. Additionally, the FA-PSO algorithm accurately estimates the building material thickness without prior knowledge of the thickness, achieving an error of less than 0.6 mm relative to the measured values. We derived the CRLB for the permittivity, conductivity, and thickness of building materials under Gaussian white noise. The CRLB were analyzed to unveil the impacts of walls thickness on the inversion accuracy algorithm performance. Furthermore, to enhance algorithmic performance, the initial particle swarm of the FA-PSO algorithm was optimised and simulation results indicate that the inversion accuracy approaches the CRLB. The proposed approach will facilitate building material selection in terms of wireless performance for civil engineer.

\end{document}